\documentclass{ifacconf}

\usepackage{graphicx}      % include this line if your document contains figures
\usepackage{natbib}        % required for bibliography
\usepackage{tikz,tikz-3dplot}
\usepackage{pgfplots} 
\usepackage{pgfgantt}
\usepackage{pdflscape}
\pgfplotsset{compat=newest} 
\pgfplotsset{plot coordinates/math parser=false}
\pgfplotsset{table/search path={figures/data}}
\usepackage{threeparttable}
\usetikzlibrary{arrows,arrows.meta}
\usepackage{subcaption}
\usepackage{import}

\usepackage{amsmath}
\allowdisplaybreaks[1]
\usepackage{amssymb}
\usepackage{wasysym}
\usepackage{mathtools}
\usepackage{xfrac}

\usepackage[english]{babel}

\usepackage{siunitx}

\newtheorem{asm}{Assumption}

\usepackage{xcolor}
\usepackage{soul}
\definecolor{highlight}{rgb}{1,1,0.8}
\sethlcolor{highlight}
\definecolor{DarkOrange}{rgb}{0.8,0.3,0}

\newcommand{\bs}[1]{\boldsymbol{#1}}
\newcommand{\mat}[1]{\mathbf{#1}}
\newcommand{\T}{^{\rm T}}
\newcommand{\abs}[1]{\left|#1\right|}
\newcommand{\norm}[1]{\left\|#1\right\|}

\newcommand*{\scale}[2][1]{\scalebox{#1}{\ensuremath{#2}}}

\newcommand{\inlinevector}[1]{\left\langle#1\right\rangle}

\newcommand{\reffig}[1]{Figure~\ref{#1}}

\begin{document}
\begin{frontmatter}

\title{\LARGE \bf Singularity-free Formation Path Following of Underactuated AUVs: Extended Version}

\author[ntnu]{Josef Matou\v{s}}
\author[ntnu]{Kristin Y. Pettersen}
\author[ntnu,padova]{Damiano Varagnolo}
\author[sintef]{Claudio Paliotta}

\address[ntnu]{Department of Engineering Cybernetics, Norwegian University of Science and Technology, Trondheim
    (name.surname@ntnu.no).}
\address[padova]{$\!$Department of Information Engineering, University of Padova, Italy.}
\address[sintef]{SINTEF Digital, Trondheim, Norway 
    (claudio.paliotta@sintef.no).}

\thanks{This work was partly supported by the Research Council of Norway through project No. 302435 and the Centres of Excellence funding scheme, project No. 223254.}
\thanks{The authors would like to thank Aurora Haraldsen for the discussions on the collision cone concept.}

    \begin{abstract}
        $\quad$This paper proposes a method for formation path following control of a fleet of underactuated autonomous underwater vehicles.
        The proposed method combines several hierarchic tasks in a null space-based behavioral algorithm to safely guide the vehicles.
        Compared to the existing literature, the algorithm includes both inter-vehicle and obstacle collision avoidance, and employs a scheme that keeps the vehicles within given operation limits.
        The algorithm is applied to a six degree-of-freedom model, using rotation matrices to describe the attitude to avoid singularities.
        Using the results of cascaded systems theory, we prove that the closed-loop system is uniformly semiglobally exponentially stable.
        We use numerical simulations to validate the results.
    \end{abstract}

        
\end{frontmatter}

    \section{Introduction}
    \vspace{-1em}
    Autonomous underwater vehicles (AUVs) are being increasingly used in a number of applications such as transportation, seafloor mapping, and other ocean energy industry-related tasks.
    It is often advantageous to perform such tasks with a group of cooperating AUVs.
    Therefore, there is a need for algorithms that can safely guide a formation of AUVs along a given path while avoiding collisions with each other and obstacles, and staying within given operation limits.

    \vspace{-0.4em}

    As presented in \cite{das_cooperative_2016}, there exists a plethora of formation path-following methods, most of them based on two concepts: coordinated path-following \citep{borhaug_2006_formation,ghabcheloo_2006_coordinated} and leader-follower \citep{rongxin_2010_leader,soorki_2011_robust}.
    In the \emph{coordinated path-following} approach, each vehicle follows a predefined path separately.
    Formation is then achieved by coordinating the motion of the vehicles along these paths.
    In this approach, the formation-keeping error (\emph{i.e.,} the difference between the actual and desired relative position of the vehicles) may initially grow as the vehicles converge to their predefined paths.
    In the \emph{leader-follower} approach, one leading vehicle follows the given path while the followers adjust their speed and position to obtain the desired formation shape.
    This latter approach tends to suffer from the lack of formation feedback due to unidirectional communication (\emph{i.e.,} the leader may not adjust its velocity based on the followers).

    %\vspace{-0.5em}

    Another formation path-following algorithmic paradigm is the so-called null-space-based behavioral (NSB) approach \citep{arrichiello_formation_2006,antonelli_experiments_2009,pang_2019_formation,eek_formation_2021}, a centralized strategy that allows to combine several hierarchic tasks.
    In the NSB framework, the control objective is expressed using multiple tasks.
    By combining these simple tasks, the vehicles exhibit the desired complex behavior.

    \vspace{-0.5em}

    This paper aims to extend our previous NSB algorithm \citep{matouvs_formation_2022} to control a fleet of AUVs.
    The previous work uses a five degree-of-freedom (5DOF) AUV model, considers only inter-vehicle collision avoidance, and proves only the stability of the path-following algorithm.
    Furthermore, the orientation of the 5DOF model was expressed using Euler angles, which causes singularities for a pitch angle of $\pm90$ degrees.
    This work applies the NSB algorithm to a full 6DOF model, uses rotation matrices to describe the attitude of the vehicles to avoid singularities, modifies and extends the tasks, and proves the stability of the combined path-following and formation-keeping tasks.
    We also add a scheme that keeps the vehicles within a given range of depths to stay within the operation limits.
    As opposed to the previous work, we do not limit the analysis to a specific low-level attitude controller.
    Consequently, the new algorithm can be integrated into existing on-board controllers.
    Assuming that the existing low-level controller allows exponential tracking, we use results from cascaded systems theory~\citep{pettersen_lyapunov_2017} to prove that the closed-loop system composed by the NSB algorithm and the low-level controller is uniformly semiglobally exponentially stable.
    We verify the results in numerical simulations.

    \vspace{-0.5em}

    The remainder of the paper is organized as follows.
    Section~\ref{sec:model} introduces the model of the AUVs.
    Section~\ref{sec:objectives} defines the formation path-following problem.
    Section~\ref{sec:control} describes the proposed modified NSB algorithm.
    The stability of the closed-loop system is proven in Section~\ref{sec:path_stability}.
    Section~\ref{sec:simulation} presents the results of the numerical simulations.
    Finally, Section~\ref{sec:conclusion} presents some concluding remarks.

    \vspace{-0.6em}
    
    \section{The AUV Model}
    \label{sec:model}
    \vspace{-1em}
%    In this section, we present the model of the AUV.
    To simplify the notation, we will denote a concatenation of vectors or scalars using angled brackets, \emph{e.g.,}
    \vspace{-0.3em}
    \begin{equation}
        \inlinevector{\mat{x}_1, \ldots, \mat{x}_N} = \left[ \mat{x}_1\T, \ldots, \mat{x}_N\T \right]\T.
    \end{equation}
    Let $\mat{p} = \inlinevector{x, y, z}$ be the position, $\mat{R} \in SO(3)$ the rotation matrix describing the orientation, $\mat{v} = \inlinevector{u, v, w}$ the linear surge, sway and heave velocities, and $\bs{\omega} = \inlinevector{p, q, r}$ the angular velocity of the vehicle.
    For brevity, let us also define the velocity vector $\bs{\nu} = \inlinevector{\mat{v}, \bs{\omega}}$.

    \vspace{-0.3em}

    Furthermore, let $\mat{V}_c = \inlinevector{V_x, V_y, V_z}$ be the velocities of an unknown, constant and irrotational ocean current, given in the inertial frame, and $\mat{v}_c = \inlinevector{u_c, v_c, w_c}$ the ocean current velocities expressed in the body-fixed coordinate frame
    \vspace{-0.1em}
    \begin{equation}
        \label{eq:body_fixed_current}
        \mat{v}_c = \mat{R}\T\,\mat{V}_c.
    \end{equation}
    \vspace{-1.9em}

    \noindent We will denote the relative linear velocities of the vehicle as $\mat{v}_r = \mat{v} - \mat{v}_c$.
    We will also denote the relative surge, sway and heave velocities as $u_r$, $v_r$ and $w_r$, and the relative velocity vector as $\bs{\nu}_r = \inlinevector{\mat{v}_r, \bs{\omega}}$.

    \vspace{-0.3em}

    Let $\mathbf{f} = \inlinevector{T_u, \bs{\delta}}$ be the vector of control inputs, where $T_u$ is the surge thrust generated by the propeller, and $\bs{\delta}$ represents the configuration of fins.
    Furthermore, let $\mat{M}$ be the mass and inertia matrix, including added mass effects, $\mat{C}(\bs{\nu}_r)$ the Coriolis centripetal matrix, also including added mass effects, and $\mat{D}(\bs{\nu}_r)$ the hydrodynamic damping matrix.
    The dynamics of the vehicle in a matrix-vector form are then \citep{fossen_handbook_2011}
    \vspace{-0.5em}
    \begin{subequations}
    \begin{align}
        \dot{\mat{p}} &= \mat{R} \mat{v}, \label{eq:p_dot}\\
        \dot{\mat{R}} &= \mat{R} \mat{S}(\bs{\omega}), \label{eq:R_dot} \\
        \mat{M} \dot{\bs{\nu}}_r
            + \bigl(\mat{C}(\bs{\nu}_r) + \mat{D}(\bs{\nu}_r)\bigr) \bs{\nu}_r + \mat{g}(\mat{R}) &= \mat{B}\mat{f}, \label{eq:nu_dot}
    \end{align} \label{eq:matrix_vector_model}
    \end{subequations}
    \vspace{-1.6em}
    
    \noindent where $\mat{g}(\mat{R})$ is the gravity and buoyancy vector, $\mat{B}$ is the actuator configuration matrix that maps the control inputs to forces and torques, and $\mat{S}: \mathbb{R}^3 \mapsto \mathfrak{so}(3)$ is the skew-symmetric matrix operator.

    \vspace*{-0.2em}

    Note that \eqref{eq:nu_dot} describes the dynamics of a generic underwater rigid body.
    In the remainder of this section, we will derive a more specific model for an AUV.
    First, let us present the necessary assumptions about the vehicle.
    \vspace{-0.3em}
    \begin{asm}
        \label{ass1}
        The vehicle is slender, torpedo-shaped, with port-starboard and top-bottom symmetry.
    \end{asm}
    \vspace{-0.6em}
    \begin{asm}
        \label{ass2}
        The hydrodynamic damping is linear.
    \end{asm}
    \vspace{-0.6em}
    \begin{asm}
        \label{ass3}
        The vehicle is neutrally buoyant, with the center of gravity (CG) and the center of buoyancy (CB) located along the same vertical axis.
    \end{asm}
    \vspace{-0.6em}
    \begin{asm}
        \label{ass4}
        The origin of the body-fixed frame is chosen such that actuators produce no sway and heave acceleration.
        In other words, there exist $f_u, t_p, t_q, t_r$ such that
        \vspace{-0.9em}
        \begin{equation}
            \mat{M}^{-1}\,\mat{B}\mat{f} = \inlinevector{f_u, 0, 0, t_p, t_q, t_r}.
            \label{eq:forces}
        \end{equation}
    \end{asm}
    \vspace{-0.65em}

    \emph{Remark.} the mechanical design of typical commercial survey AUVs satisfies Assumptions \ref{ass1} and \ref{ass3}. Assumption \ref{ass2} is valid for low-speed missions and is often used as a simplification also when designing controllers for higher-speed missions, as the higher-order damping coefficients are poorly known, and compensating for these may reduce the robustness of the control system. The general structure of $\mat{M}$, $\mat{C}(\cdot)$, $\mat{D}(\cdot)$, and $\mat{g}(\cdot)$ for vehicles that satisfy Assumptions \ref{ass1}--\ref{ass3} is shown, \emph{e.g.}, in \cite{fossen_handbook_2011}. 
    In \cite{borhaug_straight_2007}, it is shown that if a 5DOF vehicle model with port-starboard symmetry satisfies Assumptions \ref{ass2}--\ref{ass3}, the origin of the body-fixed coordinate frame can always be chosen such that Assumption~\ref{ass4} holds.
    By assuming top-bottom symmetry, the roll dynamics are decoupled from the rest of the system.
    Consequently, the procedure demonstrated in \cite{borhaug_straight_2007} can be trivially extended to 6DOFs.

    \vspace{-0.1em}

    \begin{asm}
        \label{ass5}
        The vehicle is equipped with a low-level controller that allows exponential tracking of the surge velocity, orientation, and angular velocity.
        Specifically, let $u_d, \mat{R}_d$ and $\bs{\omega}_d$ be the reference signals.
        We define an error \vspace{-1.1em}
        \begin{align}
            \widetilde{\mat{X}} &= \inlinevector{u - u_d, {\rm logm}\bigl(\widetilde{\mat{R}}\bigr), \bs{\omega} - \widetilde{\mat{R}}\T\bs{\omega}_d}, &
            \widetilde{\mat{R}} &= \mat{R}_d\T\mat{R}, \label{eq:low_level_error}
        \end{align}
        \vspace{-1.5em}
        
        \noindent where ${\rm logm} : SO(3) \mapsto \mathbb{R}^3$ is the matrix logarithm \citep{iserles_lie_2000}.
        Note that by Assumption~\ref{ass4}, $\widetilde{\mat{X}}$ is controllable through the input $\mat{f}$.
        Consider the closed-loop system \vspace{-0.25em}
        \begin{equation}
            \dot{\widetilde{\mat{X}}} = F \left( \widetilde{\mat{X}}, v, w, \mat{V}_c \right), \label{eq:low_level_closed_loop}
        \end{equation}
        \vspace{-1.6em}
        
        \noindent consisting of \eqref{eq:R_dot}, \eqref{eq:nu_dot}, and the low-level controller.
        We assume that $\widetilde{\mat{X}} = \mat{0}$ is a globally exponentially stable (GES) equilibrium of \eqref{eq:low_level_closed_loop}.
    \end{asm}

    \vspace{-0.8em}

    \emph{Remark.} The aim of this paper is to demonstrate that the proposed formation path-following algorithm can be readily implemented on vehicles with existing low-level controllers.
    Consequently, the choice of a low-level velocity and attitude controller is not discussed in this paper.
    An example of a global exponential attitude tracking controller can be found, \emph{e.g.,} in \cite{lee_global_2015}.

    \vspace*{-0.2em}

    Note that for a complete system analysis, we need to consider the underactuated sway and heave dynamics explicitly. 
    Under Assumptions \ref{ass1}--\ref{ass4}, the underactuated dynamics have the following form \vspace{-0.15em}
    \begin{subequations}
        \begin{align}
            \dot{v} &= X_v(u_r)r + Y_v(u_r)v_r + Z_v(p)w_r + \dot{v}_c, \\
            \dot{w} &= X_w(u_r)q + Y_w(u_r)w_r + Z_w(p)v_r + \dot{w}_c,
            \label{eq:underactuated_dynamics}
        \end{align}
    \end{subequations}
    \vspace{-1.6em}
    
    \noindent where $X(\cdot), Y(\cdot), Z(\cdot)$ are affine functions of the respective variables.
    From \eqref{eq:body_fixed_current}, it follows that \vspace{-0.2em}
    \begin{equation}
        \dot{\mat{v}}_c = \inlinevector{\dot{u}_c, \dot{v}_c, \dot{w}_c} = \mat{v}_c \times \bs{\omega},
    \end{equation}
    \vspace{-2em}
    
    \noindent where $\times$ denotes the vector cross product.

    \vspace{-0.9em}
    \section{Formation Path Following}
    \label{sec:objectives}
    \vspace{-1.1em}
    The goal is to control a fleet of $n$ AUVs so that they move in a prescribed formation and their barycenter follows a given path.

    \begin{figure}[b]
        %\vspace*{-2.3em}
        \centering
        \def\svgwidth{.4\textwidth}
        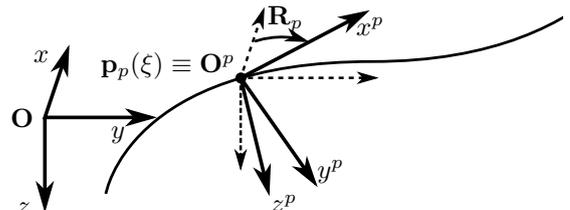
        \vspace{-1.3em}
        \caption{Definition of the path angles and path-tangential coordinate frame. $\mathbf{O}$ denotes the origin of the inertial coordinate frame, $\mathbf{O}^p$ denotes the origin of the path-tangential frame.}
        \label{fig:path}
        %\vspace{-3mm}
    \end{figure}

    \vspace{-0.4em}

    The prescribed path in the inertial coordinate frame is given by a smooth function $\mat{p}_p: \mathbb{R} \mapsto \mathbb{R}^3$.
    We assume that the path function is $\mathcal{C}^{\infty}$ and regular, \emph{i.e.,} the function is continuously differentiable and its partial derivative with respect to $\xi$ satisfies %\vspace{-0.8em}
%    \begin{equation}
$
        \norm{\frac{\partial \mat{p}_p(\xi)}{\partial \xi}} \neq 0
$.
    % \end{equation}
    % \vspace{-1.1em}
    %
    Therefore, for every point $\mat{p}_p(\xi)$ on the path, there exists a path-tangential coordinate frame $(x^p, y^p, z^p)$ and a corresponding rotation matrix $\mat{R}_p$ (see \reffig{fig:path}).

    \vspace{-0.4em}
    
    The path-following error $\mat{p}_b^p$ is given by the position of the barycenter in the path-tangential coordinate frame \vspace{-0.5em}
    \begin{align}
        \mat{p}_b^p &= \mat{R}_p\T \, \big(\mat{p}_b - \mat{p}_p(\xi)\big), &
        \mat{p}_b &= \frac{1}{n} \sum_{i=1}^n \mat{p}_i.
        \label{eq:barycenter}
    \end{align}
    \vspace{-1.2em}
    
    \noindent The goal of path following is to control the vehicles so that $\mat{p}_b^p \equiv \mat{0}_3$, where $\mat{0}_3$ is a 3-element vector of zeros.

    \begin{figure}[t]
        \centering
        \def\svgwidth{.33\textwidth}
        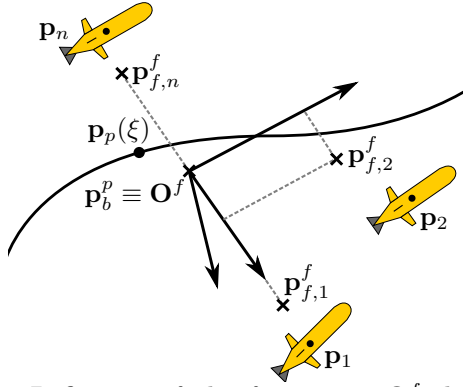
        \vspace{-1em}
        \caption{Definition of the formation. $\mathbf{O}^f$ denotes the origin of the formation-centered coordinate frame.}
        \label{fig:formation}
        \vspace{-0.3em}
    \end{figure}

    \vspace{-0.2em}

    To define the formation-keeping problem, we first define the formation-centered coordinate frame.
    This coordinate frame is created by translating the path-tangential frame into the barycenter (see \reffig{fig:formation}).
    Let $\mat{p}_{f,1}^f, \ldots, \mat{p}_{f,n}^f$ be the position vectors that represent the desired formation.
    From \reffig{fig:formation}, one can see that these vectors are constant in the formation-centered frame.
    Furthermore, the mean value of $\mat{p}_{f,i}^f$ must coincide with the barycenter.
    Since the barycenter is equivalent to the origin of the formation-centered frame, the vectors must thus satisfy \vspace{-0.35em}
    \begin{equation}
        \sum_{i=1}^n \mat{p}_{f, i}^f = \mat{0}_3. \vspace{-0.5mm}
    \end{equation}
    \vspace{-1em}

    \noindent The position of vehicle $i$ in the formation-centered frame is then given by \vspace{-0.55em}
    \begin{equation}
        \mat{p}_i^f = \mat{R}_p\T \left(\mat{p}_i - \mat{p}_b\right).
    \end{equation}  
    \vspace{-1.5em}

    \noindent The goal of formation keeping is to try to maintain $\mat{p}_i^f \equiv \mat{p}_{f, i}^f$ independently of the disturbances experienced by the agents.
    This problem can also be expressed in the inertial coordinate frame as \vspace{-0.15em}
    \begin{align}
        \mat{p}_i &\equiv \mat{R}_p \mat{p}_{f,i}^f + \mat{p}_b, &
        i &\in \left\{1, \ldots, n \right\}.
    \end{align}
    \vspace{-1.7em}

    \section{Control System}
    \label{sec:control}
    \vspace{-1em}
    The AUVs must perform the goals stated in Section \ref{sec:objectives} safely, \emph{i.e.},
    avoid collisions with other vehicles and obstacles, and remain within a given range of depths.
    An upper limit on the depth of the AUVs is needed to prevent them from colliding with the seabed or exceeding their depth rating.
    A lower limit is needed in busy environments (\emph{e.g.,} harbors), where the AUVs may otherwise collide or interfere with surface vessels.

    \vspace{-0.2em}

    To solve the formation path following problem, we propose a method that combines inter-vehicle collision avoidance (COLAV), formation keeping, line-of-sight (LOS) path following, obstacle avoidance, and depth limiting in a hierarchic manner using an NSB algorithm.
    Since the NSB algorithm outputs inertial velocity references, we also need a method for converting these to surge and orientation.

    \vspace{-0.2em}

    In this section, we first present the NSB algorithm and the associated tasks.
    We then present in Section~\ref{sec:references} a strategy for converting inertial velocity references to surge/orientation ones.

    \vspace{-0.6em}

    \subsection{NSB algorithm}
    \label{sec:NSB}
    \vspace{-0.9em}
    The NSB algorithm allows us to define and combine multiple tasks in a hierarchic manner.
    For more information, the reader is referred to \cite{antonelli_2006_kinematic}.

    \vspace{-0.4em}

    Achieving the desired behavior requires three tasks: COLAV, formation-keeping, and path-following.
    Each task will be described in detail in Sections~\ref{sec:COLAV}, \ref{sec:formation}, and \ref{sec:LOS}, while in the remainder of this subsection, we introduce some mathematical tools instrumental for describing each of these tasks.
    As we will explain in Section \ref{sec:OA}, obstacle avoidance and depth limiting will not be defined as separate tasks but rather achieved through a modification to the path-following task.
    Let us denote the variables associated with the COLAV, formation-keeping, and path-following tasks by lower indices $1$, $2$, and $3$, respectively.
    Define the so-called \emph{task variables} as $\bs{\sigma}_i = \bs{f}_i\left(\mat{p}_1, \ldots, \mat{p}_n\right), i \in \left\{1,2,3\right\}$, and their desired values as $\bs{\sigma}_{d,i}, i \in \left\{1,2,3\right\}$.

    \vspace{-0.3em}

    Furthermore, let $\bs{\upsilon}_i, i \in \left\{1,2,3\right\}$ be the desired velocities of each task.    
    In the standard NSB algorithm, $\bs{\upsilon}_i$ is obtained using the closed-loop inverse kinematics (CLIK) equation \citep{antonelli_2006_kinematic} \vspace{-0.25em}
    \begin{equation}
        \label{eq:CLIK}
        \bs{\upsilon}_i = \mat{J}_i^{\dagger}\,\bigl(\dot{\bs{\sigma}}_{d,i} - \bs{\Lambda}_i\,\widetilde{\bs{\sigma}}_i\bigr),
    \end{equation}
    where $\bs{\Lambda}_i$ is a positive definite gain matrix, $\widetilde{\bs{\sigma}}_i = \bs{\sigma}_i - \bs{\sigma}_{d, i}$, and $\mat{J}_i^{\dagger}$ is the Moore-Penrose pseudoinverse of the task Jacobian \vspace{-0.4em}
    \begin{equation}
        \mat{J}_i = \frac{\partial \bs{\sigma}_{d,i}}{\partial \inlinevector{\mat{p}_1, \ldots, \mat{p}_n}}.
    \end{equation}
    However, in our case, we need to modify this equation for each task to make it applicable to underactuated AUVs.
    %These modifications will be explained in Sections~\ref{sec:COLAV}, \ref{sec:formation}, and \ref{sec:LOS}.

    The combined desired velocity, $\bs{\upsilon}_{\rm NSB}$, is then given by \citep{antonelli_2006_kinematic}
    \vspace{-0.35em}
    \begin{equation} 
        \scale[0.95]{\bs{\upsilon}_{\rm NSB} = \bs{\upsilon}_1 + \left(\mat{I} - \mat{J}_1^{\dagger}\mat{J}_1\right)\left(\bs{\upsilon}_2 + \left(\mat{I} - \mat{J}_2^{\dagger}\mat{J}_2\right)\bs{\upsilon}_3\right),} \label{eq:NSB_with_COLAV}
    \end{equation}
    where $\mat{I}$ is an identity matrix.
    \vspace{-0.8em}

    \subsection{Inter-vehicle collision avoidance}
    \label{sec:COLAV}
    \vspace{-1em}
    Let {$d_{\rm COLAV}$} be the \emph{activation distance}, \emph{i.e.,} the distance at which the vehicles need to start performing the evasive maneuvers.
    The task variable is given by a vector of relative distances between the vehicles smaller than $d_{\rm COLAV}$
    \begin{align}
            \bs{\sigma}_1 &= \big[\norm{\mat{p}_i - \mat{p}_j}\big], &
            \begin{split} 
                \forall &i,j\in\{1,\ldots,n\}, j > i, \\
                &\norm{\mat{p}_i - \mat{p}_j} < d_{\rm COLAV}.
            \end{split}
    \end{align}
    The desired values of the task are \vspace{-0.3em}
    \begin{equation}
        \bs{\sigma}_{d,1} = d_{\rm COLAV} \, \mat{1},
    \end{equation}
    where $\mat{1}$ is a vector of ones.
    To ensure a faster response to a potential collision than in \cite{matouvs_formation_2022}, we propose the following sliding-mode-like COLAV velocity \vspace{-0.1em}
    \begin{equation}
        \bs{\upsilon}_1 = U_{\rm COLAV} \frac{\bs{\upsilon}_{1, {\rm CLIK}}}{\norm{\bs{\upsilon}_{1, {\rm CLIK}}}},
    \end{equation}
    where $U_{\rm COLAV}$ is a positive constant, $\norm{\cdot}$ is the Euclidean norm, and $\bs{\upsilon}_{1, {\rm CLIK}}$ is the velocity vector given by~\eqref{eq:CLIK}.

    \vspace{-0.2em}

    Note that this task does not guarantee robust collision avoidance.
    During the transients, the relative distance may become smaller than $d_{\rm COLAV}$.
    Therefore, to ensure collision avoidance, $d_{\rm COLAV}$ should be chosen as $d_{\rm min} + d_{\rm sec}$, where $d_{\rm min}$ is the minimum safe distance between the vehicles, and $d_{\rm sec}$ is an additional security distance.
    \vspace{-0.6em}

    \subsection{Formation keeping}
    \label{sec:formation}
    \vspace{-0.65em}
    The formation-keeping task variable is defined as
    \begin{align}
        \bs{\sigma}_2 &= \inlinevector{\bs{\sigma}_{2,1}, \ldots, \bs{\sigma}_{2,n-1}}, \label{eq:sigma_2} &
        \bs{\sigma}_{2,i} &= \mat{p}_i - \mat{p}_b,
    \end{align}
    and its desired values are \vspace{-0.1em}
    \begin{equation}
        % \bs{\sigma}_{d,2} = \begin{bmatrix}
        %     \mat{R}_p\,\mat{p}_{f,1}^p \\
        %     \vdots \\
        %     \mat{R}_p\,\mat{p}_{f,n-1}^p
        % \end{bmatrix}. \label{eq:sigma_d_2}
        \bs{\sigma}_{d,2} = \inlinevector{\mat{R}_p\,\mat{p}_{f,1}^p, \ldots, \mat{R}_p\,\mat{p}_{f,n-1}^p}  .
        \label{eq:sigma_d_2}
    \end{equation}
    Similarly to COLAV, we use the CLIK equation~\eqref{eq:CLIK} to obtain the formation-keeping velocity.
    However, as motivated in Section~\ref{sec:references}, this velocity needs to be saturated.
    The desired velocity is thus given by \vspace{-0.3em}
    \begin{equation}
        \bs{\upsilon}_2 = \mat{J}_2^{\dagger}\dot{\bs{\sigma}}_{d,2} - \upsilon_{2, \max}\mat{J}_2^{\dagger}\,{\rm sat}\left(\bs{\Lambda}_2\widetilde{\bs{\sigma}}_2\right),
        \label{eq:v_form}
    \end{equation}
    where $\upsilon_{2, \max}$ is a positive constant, and $\rm sat$ is a saturation function given by \vspace{-0.7em}
    \begin{equation}
        {\rm sat}(\mat{x}) = \mat{x} \frac{{\rm tanh}\left(\norm{\mat{x}}\right)}{\norm{\mat{x}}},
    \end{equation}
    where $\rm tanh$ is the hyperbolic tan function.
    \vspace{-0.65em}

    \subsection{Path Following}
    \label{sec:LOS}
    \vspace{-0.75em}
    Unlike the previous two tasks, the path-following task uses LOS guidance instead of CLIK.
    Let us denote the components of $\mat{p}_b^p$ as $x_b^p, y_b^p$, and $z_b^p$.
    Furthermore, let $\Delta\left(\mat{p}_b^p\right)$ be the lookahead distance of the LOS guidance law.
    Inspired by \cite{belleter_2019_observer}, we choose an error-dependent lookahead distance \vspace{-0.2em}
    \begin{equation}
        \Delta\left(\mat{p}_b^p\right) = \sqrt{\Delta_0^2 + \left(x_b^p\right)^2 + \left(y_b^p\right)^2 + \left(z_b^p\right)^2}
        \label{eq:delta}
    \end{equation}
    where $\Delta_0$ is a positive constant.
    The LOS velocity then is \vspace{-0.8em}
    \begin{equation}
        \bs{\upsilon}_{\rm LOS} = \mat{R}_p\,\scale[0.75]{\bigg\langle}\Delta\left(\mat{p}_b^p\right), -y_b^p, -z_b^p\scale[0.75]{\bigg\rangle}\,\frac{U_{\rm LOS}}{D},
        \label{eq:v_LOS}
    \end{equation}
    where $U_{\rm LOS} > 0$ is the desired path-following speed, and \vspace{-0.2em}
    \begin{equation}
        D = \sqrt{\Delta(\cdot)^2 + \left(y_b^p\right)^2 + \left(z_b^p\right)^2}.
    \end{equation}
    The task velocity is then given by \vspace{-0.2em}
    \begin{equation}
        \bs{\upsilon}_3 = \mat{1}_n \otimes \bs{\upsilon}_{\rm LOS}
        \label{eq:v_3}
    \end{equation}
    where $\otimes$ is the Kronecker tensor product.

    \vspace{-0.2em}
    
    Note that the path parameter $\xi$ in~\eqref{eq:barycenter} can be treated as an additional degree of freedom in the control design, and used to get a stable behavior of the along-track error $x_b^p$. 
    Inspired by \cite{belleter_2019_observer}, we choose the update law of $\xi$ as
    \vspace{-0.7em}
    \begin{equation}
        \dot{\xi} = \norm{\frac{\partial \mat{p}_p(\xi)}{\partial \xi}}^{-1} U_{\rm LOS} \left(\frac{\Delta}{D} + k_{\xi}\,\frac{x_b^p}{\sqrt{1+\left(x_b^p\right)^2}}\right),\label{eq:path_update}
    \end{equation}
    where $k_{\xi}$ is a positive gain.
    % Damiano: maybe we should also say something along "other choices could have been possible, e.g., by taking inspiration from X Y Z, but our experience seems to indicate that the final performance is not too much affected by this choice"? 
    \vspace{-0.65em}

    \subsection{Obstacle avoidance and depth limiting}
    \label{sec:OA}
    \vspace{-0.75em}
    Obstacle avoidance is typically implemented individually for each vehicle \citep{antonelli_2006_kinematic}.
    However, we propose to perform this task globally by incorporating it into the path-following algorithm so that it does not interfere with the inter-vehicle COLAV. %(indeed, inter-vehicle COLAV may be ensured by changing opportunely the relative velocities of the vehicles).

    \vspace{-0.3em}

\begin{figure}[t]
        \centering
        \begin{subfigure}[t]{0.23\textwidth}
            \centering
            \def\svgwidth{\textwidth}
            %% Creator: Inkscape 1.2.1 (9c6d41e410, 2022-07-14), www.inkscape.org
%% PDF/EPS/PS + LaTeX output extension by Johan Engelen, 2010
%% Accompanies image file 'obstacle_avoidance.pdf' (pdf, eps, ps)
%%
%% To include the image in your LaTeX document, write
%%   \input{<filename>.pdf_tex}
%%  instead of
%%   \includegraphics{<filename>.pdf}
%% To scale the image, write
%%   \def\svgwidth{<desired width>}
%%   \input{<filename>.pdf_tex}
%%  instead of
%%   \includegraphics[width=<desired width>]{<filename>.pdf}
%%
%% Images with a different path to the parent latex file can
%% be accessed with the `import' package (which may need to be
%% installed) using
%%   \usepackage{import}
%% in the preamble, and then including the image with
%%   \import{<path to file>}{<filename>.pdf_tex}
%% Alternatively, one can specify
%%   \graphicspath{{<path to file>/}}
%% 
%% For more information, please see info/svg-inkscape on CTAN:
%%   http://tug.ctan.org/tex-archive/info/svg-inkscape
%%
\begingroup%
  \makeatletter%
  \providecommand\color[2][]{%
    \errmessage{(Inkscape) Color is used for the text in Inkscape, but the package 'color.sty' is not loaded}%
    \renewcommand\color[2][]{}%
  }%
  \providecommand\transparent[1]{%
    \errmessage{(Inkscape) Transparency is used (non-zero) for the text in Inkscape, but the package 'transparent.sty' is not loaded}%
    \renewcommand\transparent[1]{}%
  }%
  \providecommand\rotatebox[2]{#2}%
  \newcommand*\fsize{\dimexpr\f@size pt\relax}%
  \newcommand*\lineheight[1]{\fontsize{\fsize}{#1\fsize}\selectfont}%
  \ifx\svgwidth\undefined%
    \setlength{\unitlength}{74.46179608bp}%
    \ifx\svgscale\undefined%
      \relax%
    \else%
      \setlength{\unitlength}{\unitlength * \real{\svgscale}}%
    \fi%
  \else%
    \setlength{\unitlength}{\svgwidth}%
  \fi%
  \global\let\svgwidth\undefined%
  \global\let\svgscale\undefined%
  \makeatother%
  \begin{picture}(1,1.03736494)%
    \lineheight{1}%
    \setlength\tabcolsep{0pt}%
    \put(0,0){\includegraphics[width=\unitlength,page=1]{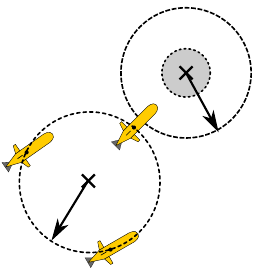}}%
    \put(0.80938669,0.64433968){\color[rgb]{0,0,0}\makebox(0,0)[lt]{\lineheight{1.25}\smash{\begin{tabular}[t]{l}$r_o$\end{tabular}}}}%
    \put(0.37012909,0.29928039){\color[rgb]{0,0,0}\makebox(0,0)[lt]{\lineheight{1.25}\smash{\begin{tabular}[t]{l}$\mathbf{p}_b$\end{tabular}}}}%
    \put(0.61858689,0.86345774){\color[rgb]{0,0,0}\makebox(0,0)[lt]{\lineheight{1.25}\smash{\begin{tabular}[t]{l}$\mathbf{p}_o$\end{tabular}}}}%
    \put(0.15702348,0.23254079){\color[rgb]{0,0,0}\makebox(0,0)[lt]{\lineheight{1.25}\smash{\begin{tabular}[t]{l}$r_f$\end{tabular}}}}%
    \put(0.48454677,0.42015374){\color[rgb]{0,0,0}\makebox(0,0)[lt]{\lineheight{1.25}\smash{\begin{tabular}[t]{l}$\mathbf{p}_1$\end{tabular}}}}%
    \put(0.3571906,0.12301686){\color[rgb]{0,0,0}\makebox(0,0)[lt]{\lineheight{1.25}\smash{\begin{tabular}[t]{l}$\mathbf{p}_2$\end{tabular}}}}%
    \put(0.02594282,0.51152543){\color[rgb]{0,0,0}\makebox(0,0)[lt]{\lineheight{1.25}\smash{\begin{tabular}[t]{l}$\mathbf{p}_3$\end{tabular}}}}%
  \end{picture}%
\endgroup%

            \vspace{-1.5em}
            \caption{Obstacle and formation radii}
            \label{fig:obstacle_radius}
            \vspace{-0.7em}
        \end{subfigure}   
        \begin{subfigure}[t]{0.23\textwidth}
            \centering
            \def\svgwidth{\textwidth}
            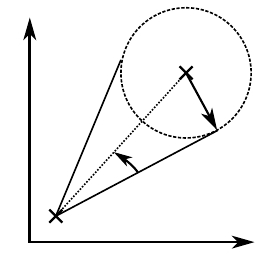
            \vspace{-1.5em}
            \caption{Collision cone}
            \label{fig:collision_cone}
            \vspace{-0.7em}
        \end{subfigure} 
        \caption{Illustration motivating the obstacle avoidance constraint~\eqref{eq:obstacle_avoidance_condition} and conflict condition~\eqref{eq:collision-conflict}.}     
        \vspace{-0.3em}
    \end{figure}

    To arrive at the proposed algorithm, we first restrict the obstacle avoidance maneuvers to the $xy$-plane to avoid interfering with the subsequent depth-limiting logic.
    Let $\mat{p}_o = \inlinevector{x_o, y_o, z_o}$ be the position of the obstacle and $r_o$ the obstacle avoidance radius.
    Note that $r_o$ must be chosen sufficiently large to cover the size of both the obstacle and the AUV. %(and relative uncertainties).
    Furthermore, let us define the formation radius
    $
        r_f = \max_{i \in \{1, \ldots, n\}} \norm{\inlinevector{x_b-x_i, y_b-y_i}}
    $
    and the relative position $\mat{p}_{\rm rel} = \inlinevector{x_o - x_b, y_o - y_b}$. As illustrated in \reffig{fig:obstacle_radius}, obstacle avoidance is ensured if \vspace{-0.8em}
    \begin{equation}
        \norm{\mat{p}_{\rm rel}} \geq r_o + r_f. \label{eq:obstacle_avoidance_condition}
    \end{equation}

    \vspace{-0.5em}

    To guarantee obstacle avoidance, we utilize the collision cone concept \citep{chakravarthy_obstacle_1998}. 
    Inspired by \cite{wiig_collision_2019}, we employ a constant avoidance angle and define a switching condition.
    More precisely, let \vspace{-1.1em}
    \begin{equation}
        \bs{\upsilon}_{\rm rel} = \inlinevector{\dot{x}_{\rm LOS} - \dot{x}_o, \dot{y}_{\rm LOS} - \dot{y}_o}
    \end{equation}
    denote the relative line-of-sight velocity ($\dot{x}_{\rm LOS}$ and $\dot{y}_{\rm LOS}$ are the components of $\bs{\upsilon}_{\rm LOS}$).
    As shown in~\reffig{fig:collision_cone}, a conflict between the AUVs and obstacle arises if \vspace{-0.25em}
    \begin{align}
        \abs{\angle\left(\mat{p}_{\rm rel}, \bs{\upsilon}_{\rm rel}\right)} &\leq \alpha, &
        \alpha &= \sin^{-1}\left(\frac{r_o+r_f}{\norm{\mat{p}_{\rm rel}}}\right),
        \label{eq:collision-conflict}
    \end{align}
    where $\angle\left(\mat{a}, \mat{b}\right)$ denotes the angle between two vectors.
    \vspace{-0.15em}

    The obstacle avoidance task is activated if simultaneously such a conflict arises and the cone angle satisfies 
    $\alpha \geq \alpha_{\min}$, where $0 < \alpha_{\min} \ll \pi / 2$.
    Note that \cite{wiig_collision_2019} use a switching condition based on distance, \emph{i.e.,}
    $\norm{\mat{p}_{\rm rel}} \leq d_{\min}$.
    Since our definition of a safe distance \eqref{eq:obstacle_avoidance_condition} is not constant, we instead suggest using a switching rule based on the cone angle.

    \vspace{-0.15em}

    When the task is active, the $x$- and $y$-components of the LOS velocity are replaced by the obstacle avoidance velocity $\bs{\upsilon}_{\rm OA}$ given by \vspace{-0.15em}
    \begin{align}
        \bs{\upsilon}_{\rm OA} &= \norm{\bs{\upsilon}_{\rm rel}} \inlinevector{\cos(\psi_{\rm OA}), \sin(\psi_{\rm OA})} + \inlinevector{\dot{x}_o, \dot{y}_o}, \\
        \psi_{\rm OA} &= {\rm atan}_2 \left(y_o - y_b, x_o - x_b\right) \pm \alpha,
    \end{align}
    where ${\rm atan}_2$ is the four-quadrant inverse tan.
    Note that $\psi_{\rm OA}$ has two solutions corresponding to the clockwise and counterclockwise directions.
    Inspired by \cite{haraldsen_reactive_2021}, we propose the following method for choosing a direction:
    When the conflict first happens, we choose the value of $\psi_{\rm OA}$ that is closer to the direction of $\bs{\upsilon}_{\rm rel}$.
    Afterwards, we maintain the same direction.

    \vspace{-0.15em}

    As for the depth-limiting logic,
    let $z_{\min}$ and $z_{\max}$ be the operation limits.
    We assume the limits to be wide enough to accommodate the formation.
    We then propose to replace the $z$-component of the LOS velocity with a depth-limiting velocity $\dot{z}_{\rm lim}$ given by \vspace{-0.15em}
    \begin{equation}
        \dot{z}_{\rm lim} = \begin{cases}
            \upsilon_z, & \text{if } \min_{i \in \{1, \ldots, n\}} z_i \leq z_{\min}, \\
            -\upsilon_z, & \text{if } \max_{i \in \{1, \ldots, n\}} z_i \geq z_{\max}, \\
            \dot{z}_{\rm LOS}, & \text{otherwise},
        \end{cases}
    \end{equation}
    where $\upsilon_z$ is a positive constant.

    \vspace{-0.9em}
    \subsection{Surge and orientation references}
    \label{sec:references}
    \vspace{-0.9em}
    Since the NSB algorithm outputs inertial velocity references, we also need a method for converting these to surge and orientation references.
    The strategy for choosing these references changes depending on whether the avoidance or depth-limiting tasks are active.
    The proposed strategy allows us to prove the closed-loop stability of both the path-following and formation-keeping tasks (\emph{c.f.} \cite{arrichiello_formation_2006}, where no stability proofs are given, and \cite{eek_formation_2021,matouvs_formation_2022}, that only prove the stability of the path-following task).

    \vspace{-0.25em}

    First, let us consider the case when neither the avoidance nor depth-limiting tasks are active.
    Note that due to the properties of the task velocities and Jacobians, \eqref{eq:NSB_with_COLAV} can be simplified to \vspace{-0.35em}
    \begin{equation}
        \bs{\upsilon}_{\rm NSB} = \bs{\upsilon}_2 + \bs{\upsilon}_3.
        \label{eq:NSB_nominal_simplified}
    \end{equation}
    \vspace{-2em}

    \noindent Let $\bs{\upsilon}_{{\rm NSB}, i}$ denote the desired velocity of vehicle $i$.    
    To achieve the desired behavior, the surge reference $u_{d, i}$ must satisfy \vspace{-0.8em}
    \begin{equation}
        u_{d, i} = \sqrt{\norm{\bs{\upsilon}_{{\rm NSB}, i}}^2 - v_i^2 - w_i^2},
        \label{eq:surge_reference}
    \end{equation}
    \vspace{-1.7em}

    \noindent However, \eqref{eq:surge_reference} can only be satisfied if \vspace{-0.35em}
    \begin{equation}
        \norm{\bs{\upsilon}_{{\rm NSB}, i}}^2 \geq v_i^2 + w_i^2.
    \end{equation}
    In addition, AUVs typically need to maintain a minimum surge velocity to be able to maneuver, implying a stricter inequality \vspace{-0.15em}
    \begin{equation}
        \norm{\bs{\upsilon}_{{\rm NSB}, i}}^2 \geq u_{\min}^2 + v_i^2 + w_i^2
        \label{eq:NSB_speed_condition}
    \end{equation}
    where $u_{\min} > 0$.
    This inequality can be satisfied by choosing a time-varying path-following speed $U_{\rm LOS}$.

    \vspace{-0.25em}

    Substituting task velocity definitions \eqref{eq:v_form} and \eqref{eq:v_3} into \eqref{eq:NSB_nominal_simplified} and exploiting the structure of the task Jacobian $\mat{J}_2$, we get that the NSB velocity of vehicle $i$ is given by \vspace{-0.3em}
    \begin{equation}
        \bs{\upsilon}_{{\rm NSB}, i} = \bs{\upsilon}_{\rm LOS} + \dot{\mat{R}}_p(\xi)\mat{p}_{f,i}^f + \bs{\upsilon}_{2,i}, \label{eq:v_NSB_i}
        \vspace*{-0.55em}
    \end{equation}
    where \vspace{-0.45em}
    \begin{equation}
        \inlinevector{\bs{\upsilon}_{2, 1}, \ldots, \bs{\upsilon}_{2, n}} = 
        - \upsilon_{2, \max}\,{\rm sat}\left(\mat{J}_2^{\dagger}\bs{\Lambda}_2 \widetilde{\bs{\sigma}}_2\right). \label{eq:v_form_components}
    \end{equation}
    Let $\bs{\omega}_p(\xi)$ be a vector such that \vspace{-0.35em}
    \begin{equation}
        \dot{\mat{R}}_p(\xi) = \mat{R}_p(\xi)\,\mat{S}\bigl(\bs{\omega}_p(\xi)\bigr)\,\dot{\xi}. \label{eq:R_p_dot}
    \end{equation}
    \eqref{eq:path_update} implies the following upper bound \vspace{-0.35em}
    \begin{equation}
        \abs{\dot{\xi}} \leq \norm{\frac{\partial \mat{p}_p(\xi)}{\partial \xi}}^{-1} U_{\rm LOS} \left(1 + k_{\xi}\right). \label{eq:path_update_bound}
    \end{equation}
    Substituting \eqref{eq:v_form_components}, \eqref{eq:R_p_dot}, and \eqref{eq:path_update_bound} into \eqref{eq:v_NSB_i}, we get the following lower bound on the NSB velocity \vspace{-0.35em}
    \begin{align}
        \norm{\bs{\upsilon}_{{\rm NSB}, i}} &\geq U_{\rm LOS} \left(1
            - \norm{\bs{\omega}_p}\norm{\scale[1]{\frac{\partial \mat{p}_p}{\partial \xi}}}^{-1}\norm{\mat{p}_{f, i}^f} \left(1 + k_{\xi}\right)\right) \nonumber \\
            &\quad - \upsilon_{2, \max}.
    \end{align}
    Now, assuming the existence of an upper bound on the product $\norm{\bs{\omega}_p(\xi)}\norm{\partial \mat{p}_p(\xi) / \partial \xi}^{-1}$, there exists a positive constant $k_{\rm NSB}$ such that for every vehicle \vspace{-0.15em}
    \begin{equation}
        \norm{\bs{\upsilon}_{{\rm NSB}, i}} \geq (1 - k_{\rm NSB})U_{\rm LOS} - \upsilon_{2, \max}.
    \end{equation}
    Assuming that $k_{\rm NSB} < 1$, we can satisfy \eqref{eq:NSB_speed_condition} by choosing \vspace{-0.15em}
    \begin{equation}
        U_{\rm LOS} = \frac{\upsilon_{2, \max} + \max_i \sqrt{v_i^2 + w_i^2 + u_{\min}^2}}{1 - k_{\rm NSB}}.
    \end{equation}
    However, the $\max$ function would introduce switching behavior.
    To avoid this, we approximate the former with \vspace{-0.35em}
    \begin{equation}
        U_{\rm LOS} = \frac{\upsilon_{2, \max} + \sqrt{\sum_{i=1}^n \left(v_i^2 + w_i^2\right) + u_{\min}^2}}{1 - k_{\rm NSB}}.
        \label{eq:U_LOS}
    \end{equation}
    \vspace{-1.2em}

    If the avoidance or depth-limiting tasks are active, we still choose $U_{\rm LOS}$ in accordance with \eqref{eq:U_LOS}.
    However, since \eqref{eq:NSB_speed_condition} cannot be satisfied with a generic NSB velocity \eqref{eq:NSB_with_COLAV}, we choose the surge reference as \vspace{-0.25em}
    \begin{equation}
        u_{d, i} = 
        \begin{cases}
            \sqrt{\norm{\bs{\upsilon}_{{\rm NSB}, i}}^2 - v_i^2 - w_i^2},  &\text{if \eqref{eq:NSB_speed_condition} satisfied,} \\
            u_{\rm min}, &\text{otherwise}.
        \end{cases}
    \end{equation}
    \vspace{-1em}

    Finally, let us discuss the choice of desired orientation.
    Let $\overline{\bs{\upsilon}}_{{\rm NSB}, i}$ and $\overline{\mat{v}}_{i}$ denote normalized vectors.
    We are seeking $\mat{R}_{d, i} \in SO(3)$ such that \vspace{-0.25em}
    \begin{equation}
        \overline{\bs{\upsilon}}_{{\rm NSB}, i} = \mat{R}_{d, i}\,\overline{\mat{v}}_{i}.
        \label{eq:rotation_reference}
    \end{equation}
    Assume that at a given time, there is $\mat{R}_{d, i}$ that satisfies \eqref{eq:rotation_reference}.
    Differentiating \eqref{eq:rotation_reference} with respect to time yields \vspace{-0.25em}
    \begin{equation}
        \dot{\overline{\bs{\upsilon}}}_{{\rm NSB}, i} = \mat{R}_{d, i}\,\mat{S}(\bs{\omega}_{d, i})\,\overline{\mat{v}}_{i} + \mat{R}_{d, i}\,\dot{\overline{\mat{v}}}_{i},
        \label{eq:rotation_reference_derivative1}
    \end{equation}
    where $\bs{\omega}_{d, i}$ is the desired angular velocity of the vehicle.     
    Let us define\vspace{-0.35em}
    \begin{align}
        \bs{\omega}_{\bs{\upsilon}_{{\rm NSB}, i}} &= \overline{\bs{\upsilon}}_{{\rm NSB}, i} \times \dot{\overline{\bs{\upsilon}}}_{{\rm NSB}, i}, &
        \bs{\omega}_{\mat{v}_{i}} &= \overline{\mat{v}}_{i} \times \dot{\overline{\mat{v}}}_{i}.
        \label{eq:omega_v}
    \end{align}    
    Then, \eqref{eq:rotation_reference_derivative1} can be rewritten as \vspace{-0.35em}
    \begin{equation}
        \bs{\omega}_{\bs{\upsilon}_{{\rm NSB}, i}} \times \overline{\bs{\upsilon}}_{{\rm NSB}, i} = 
        \mat{R}_{d, i} \left( \bs{\omega}_{d, i} \times \overline{\mat{v}}_{i} + \bs{\omega}_{\mat{v}_{i}} \times \overline{\mat{v}}_{i} \right).
    \end{equation}
    Therefore, the desired angular velocity must satisfy \vspace{-0.15em}
    \begin{equation}
        \left( \bs{\omega}_{d, i} + \bs{\omega}_{\mat{v}_{i}} 
        - \mat{R}_{d, i}\T \bs{\omega}_{\bs{\upsilon}_{{\rm NSB}, i}} \right) \times \overline{\mat{v}}_{i} = \mat{0}. \label{eq:omega_ref_equation}
    \end{equation}
    Thus, instead of finding $\mat{R}_{d, i}$ directly, we propose to choose \vspace{-0.95em}
    \begin{equation}
        \bs{\omega}_{d, i} = \mat{R}_{d, i}\T \bs{\omega}_{\bs{\upsilon}_{{\rm NSB}, i}} - \bs{\omega}_{\mat{v}_{i}},
        \label{eq:omega_ref}
    \end{equation}
    and then evolve the desired orientation according to \vspace{-0.35em}
    \begin{equation}
        \dot{\mat{R}}_{d, i} = \mat{R}_{d, i} \mat{S}(\bs{\omega}_{d, i}).
    \end{equation}
    Note that choosing $\bs{\omega}_{d, i}$ according to \eqref{eq:omega_ref} leads to the smallest (in terms of Euclidean norm) angular velocity that satisfies \eqref{eq:omega_ref_equation}.
    We also note that there exists a subspace of angular velocities that satisfy \eqref{eq:omega_ref_equation} and a subspace of rotation matrices that satisfy \eqref{eq:rotation_reference}.
    This differs from three degree-of-freedom (3DOF) \citep{eek_formation_2021,arrichiello_formation_2006} and 5DOF \citep{matouvs_formation_2022} models, for which only one solution exists.

    \vspace{-0.7em}
    \section{Closed-Loop Analysis}
    \label{sec:path_stability}
    \vspace{-1.2em}
    In this section, we analyze the closed-loop behavior of the system.
    Throughout this section, we assume that neither the avoidance nor depth-limiting tasks are active.
    Let us define the combined formation-keeping and path-following error as \vspace{-0.7em}
    \begin{equation}
        \widetilde{\bs{\sigma}} = \inlinevector{\widetilde{\bs{\sigma}}_2, \left(\mat{p}_b^p\right)},
    \end{equation}
    and the combined low-level controller error as
    \begin{equation}
        \widetilde{\mat{X}} = \inlinevector{\widetilde{\mat{X}}_1, \ldots, \widetilde{\mat{X}}_n}.
    \end{equation}
    First, let us investigate the closed-loop dynamics of $\widetilde{\bs{\sigma}}$.
    Differentiating \eqref{eq:sigma_2}, \eqref{eq:sigma_d_2}, and \eqref{eq:barycenter} with respect to time yields \vspace{-0.35em}
    \begin{subequations}
    \begin{align}
        \dot{\widetilde{\bs{\sigma}}}_2 &= \mat{J}_2\dot{\mat{p}} - \dot{\bs{\sigma}}_{d, 2}, \qquad \dot{\mat{p}} = \inlinevector{\dot{\mat{p}}_1, \ldots, \dot{\mat{p}}_n} \\
        \dot{\mat{p}}_b^p &= \mat{R}_p\T\left(\frac{1}{n}\sum_{i=1}^n\dot{\mat{p}}_i - \dot{\mat{p}}_p\right) - \mat{S}\big(\bs{\omega}_p\dot{\xi}\big)\mat{p}_b^p.
    \end{align} \label{eq:dot_sigma_tilde}
    \end{subequations}
    From \eqref{eq:p_dot} and \eqref{eq:low_level_error} it follows that $\dot{\mat{p}}_i$ is given by \vspace{-0.25em}
    \begin{align}
        \dot{\mat{p}}_i &= \mat{R}_i\mat{v}_i = {\rm expm}\left(\bs{\delta}_i\right)\mat{R}_{d, i}\inlinevector{u_{d, i} + \widetilde{u}_i, v_i, w_i}, \label{eq:p_dot_closed_loop}
    \end{align}
    with \vspace{-0.55em}
    \begin{align}
        {\rm expm}(\bs{\delta}) &= \cos\theta\,\mat{I} + s\mat{S}(\bs{\delta}) + c\mat{S}(\bs{\delta})^2\mathrlap{,} &
        &\begin{array}{l}
        \scale[1]{\theta = \norm{\bs{\delta}},} \\
        \scale[1]{s = \frac{\sin(\theta)}{\theta},} \\
        \scale[1]{c = \frac{1 - \cos(\theta)}{\theta^2}.}
        \end{array} \label{eq:expm}
    \end{align}
    Substituting \eqref{eq:expm}, \eqref{eq:surge_reference}, and \eqref{eq:rotation_reference} into \eqref{eq:p_dot_closed_loop} we get \vspace{-0.25em}
    \begin{equation}
        \begin{split}
        \dot{\mat{p}}_i &= \bs{\upsilon}_{{\rm NSB}, i} 
                    + s(\bs{\delta}_i \times \bs{\upsilon}_{{\rm NSB}, i}) \\
                    & \quad + c\,\bs{\delta}_i\times(\bs{\delta}_i \times \bs{\upsilon}_{{\rm NSB}, i})
                    + \mat{R}_{i}\inlinevector{\widetilde{u}_i, 0, 0}.
        \end{split} \label{eq:p_dot_closed_loop_2}
    \end{equation}
    Defining a perturbing term $\mat{g}_i$ as \vspace{-0.35em}
    \begin{equation}
        \mat{g}_i = s(\bs{\delta}_i \times \bs{\upsilon}_{{\rm NSB}, i}) 
                    + c\,\bs{\delta}_i\times(\bs{\delta}_i \times \bs{\upsilon}_{{\rm NSB}, i}) 
                    + \mat{R}_{i}\inlinevector{\widetilde{u}_i, 0, 0},
                    \label{eq:g_i}
    \end{equation}
    and substituting \eqref{eq:g_i} and \eqref{eq:p_dot_closed_loop_2} into \eqref{eq:dot_sigma_tilde} yields \vspace{-0.2em}
    \begin{subequations}
        \begin{align}
            \dot{\widetilde{\bs{\sigma}}}_2 &= \mat{J}_2\bs{\upsilon}_{\rm NSB} - \dot{\bs{\sigma}}_{d, 2}
                                            +\mat{J}_2\mat{G}, \qquad \mat{G} = \inlinevector{\mat{g}_1, \ldots, \mat{g}_n} \\
            \dot{\mat{p}}_b^p &= \mat{R}_p\T\left(\!\frac{1}{n}\sum_{i=1}^n\left(\bs{\upsilon}_{{\rm NSB}, i} + \mat{g}_i\right) - \dot{\mat{p}}_{\mathrlap{p}}\right) - \mat{S}\big(\bs{\omega}_p\dot{\xi}\big)\mat{p}_b^p \mathrlap{.}
        \end{align} \label{eq:outer_loop}
    \end{subequations}
    Now, to account for the underactuated dynamics, we define a vector of concatenated sway and heave velocities as \vspace{-0.3em}
    \begin{align}
        \mat{v}_u &= \inlinevector{v_1, w_1, \ldots, v_n, w_n}, &
        \mat{v}_{u, c} &= \mat{1}_n \otimes \inlinevector{v_c, w_c}.
    \end{align}
    The underactuated dynamics can then be written as
    \begin{equation}
        \dot{\mat{v}}_u = \mat{X}\bs{\Omega} + \mat{Y}\left(\mat{v}_u - \mat{v}_{u, c}\right) + \dot{\mat{v}}_{u, c}, \label{eq:underactuated_dynamics_closed_loop}
    \end{equation}
    where $\bs{\Omega} = \inlinevector{\bs{\omega}_1, \ldots, \bs{\omega}_n}$, and $\mat{X}$ and $\mat{Y}$ are block diagonal matrices consisting of blocks $\mat{X}_1, \ldots, \mat{X}$ and $\mat{Y}_1, \ldots, \mat{Y}_n$, that are given by \vspace{-0.25em}
    \begin{align}
        \mat{X}_i &= 
        \begin{bmatrix}
            0 & 0 & X_v(u_{r, i}) \\
            0 & X_w(u_{r, i}) & 0
        \end{bmatrix}, \\
        \mat{Y}_i &=
        \begin{bmatrix}
            Y_v(u_{r,i}) & Z_v(p_i) \\
            Z_w(p_i) & Y_w(u_{r, i})
        \end{bmatrix}.
    \end{align}
    \vspace{-0.65em}

    \begin{thm} \label{thm1}
        Let Assumptions \ref{ass1}--\ref{ass5} be satisfied.
        Then, $\bigl\langle\widetilde{\bs{\sigma}}, \widetilde{\bs{X}}\bigr\rangle = \mat{0}$ is a uniformly semiglobally exponentially stable (USGES) equilibrium point of the closed-loop system \eqref{eq:outer_loop}, \eqref{eq:low_level_closed_loop}, \eqref{eq:underactuated_dynamics_closed_loop}.
        Moreover, if the second and third partial derivatives of $\mat{p}_p(\xi)$ with respect to $\xi$ are bounded and \eqref{eq:boudedness_condition} is satisfied, the underactuated sway and heave dynamics are bounded near the manifold $\bigl\langle\widetilde{\bs{\sigma}}, \widetilde{\bs{X}}\bigr\rangle = \mat{0}$.
    \end{thm}
    \vspace{-0.9em}
    \begin{pf}
        We analyze the closed-loop system as a cascade where $\widetilde{\bs{X}}$ perturbs the dynamics of $\widetilde{\bs{\sigma}}$ through $\mat{G}$.
        Consider the nominal dynamics of $\widetilde{\bs{\sigma}}$ (\emph{i.e.}, \eqref{eq:outer_loop} with $\mat{G} = \mat{0}$) and the following Lyapunov function candidate \vspace{-0.25em}
        \begin{equation}
            V = \frac{1}{2}\,\widetilde{\bs{\sigma}}\T\widetilde{\bs{\sigma}} = \frac{1}{2}\left(\widetilde{\bs{\sigma}}_2\T\widetilde{\bs{\sigma}}_2 + \left(\mat{p}_b^p\right)\T\mat{p}_b^p\right).
        \end{equation}
        The time-derivative of $V$ is \vspace{-0.25em}
        \begin{equation}
            \begin{split}
                \dot{V} &= \widetilde{\bs{\sigma}}_2\T\left(\mat{J}_2\bs{\upsilon}_{\rm NSB} - \dot{\bs{\sigma}}_{d, 2}\right)
                - \left(\mat{p}_b^p\right)\T\mat{S}\left(\bs{\omega}_p\dot{\xi}\right)\mat{p}_b^p \\ 
                &\quad + \left(\mat{p}_b^p\right)\T\mat{R}_p\T\bigg(\frac{1}{n}\sum_{i=1}^n\bs{\upsilon}_{{\rm NSB}, i} - \dot{p}_p\bigg) .
            \end{split} 
            \label{eq:Lyapunov_analysis_step1}
        \end{equation}
        Due to the properties of the NSB tasks defined in Sections \ref{sec:formation} and \ref{sec:LOS}, the following identities hold: \vspace{-0.45em}
        \begin{align}
            \mat{J}_2\bs{\upsilon}_{\rm NSB} &= \mat{J}_2\bs{\upsilon}_2, &
            \sum_{i=1}^n \bs{\upsilon}_{{\rm NSB}, i} &= \bs{\upsilon}_{\rm LOS}.
        \end{align}
        By definition (see Section \ref{sec:objectives}), $\mat{R}_p$ must satisfy \vspace{-0.35em}
        \begin{align}
            \mat{R}_p\T\dot{p}_p &= \dot{\xi}\norm{\scale[1]{\frac{\partial \mat{p}_p(\xi)}{\partial \xi}}}^{-1}\mat{e}_1, &
            \mat{e}_1 &= \inlinevector{1, 0, 0}.
        \end{align}
        Substituting \eqref{eq:v_form} and \eqref{eq:v_LOS} into \eqref{eq:Lyapunov_analysis_step1} leads to \vspace{-0.25em}
        \begin{equation}
            \begin{split}
            \dot{V} &= -\upsilon_{2, \rm max} \widetilde{\bs{\sigma}}_2\T {\rm sat}\left(\bs{\Lambda}_2\widetilde{\bs{\sigma}}_2\right) \\
                    &\quad -U_{\rm LOS} \scale[1]{\left(
                        k_{\xi}\frac{\left(x_b^p\right)^2}{\sqrt{1 + \left(x_b^p\right)^2}} +
                        \frac{\left(y_b^p\right)^2}{D} +
                        \frac{\left(z_b^p\right)^2}{D}
                      \right)}.
            \end{split}
            \label{eq:Lyapunov_analysis_step2}
        \end{equation}
        For any $\widetilde{\bs{\sigma}} \in \left\{\widetilde{\bs{\sigma}} \in \mathbb{R}^{3n}: \norm{\widetilde{\bs{\sigma}}} \leq r\right\}$, the following holds: \vspace{-0.25em}
        \begin{equation}
            \begin{split}
            \dot{V} &\leq  - \upsilon_{2, \max}\lambda_{2, \min}\scale[1]{\frac{{\rm tanh}(r)}{r}}\norm{\widetilde{\bs{\sigma}}_2}^2 \\
                    &\quad - U_{\rm LOS}\min\left\{\scale[1]{\frac{k_{\xi}}{\sqrt{1+r^2}}, \frac{1}{\sqrt{\Delta_0^2+2r^2}}}\right\}\norm{\mat{p}_b^p}^2,
            \end{split}
            \label{eq:Lyapunov_analysis_step3}
        \end{equation}
        where $\lambda_{2, \min}$ is the smallest eigenvalue of $\bs{\Lambda}_2$.
        From \eqref{eq:Lyapunov_analysis_step3}, we conclude that the derivative of $V$ satisfies
        \begin{equation}
            \dot{V} \leq - k_r \norm{\widetilde{\bs{\sigma}}}^2,
        \end{equation}
        where
        \begin{equation}
            k_r = \min \left\{\upsilon_{2, \max}\lambda_{2, \min}\scale[1]{\frac{{\rm tanh}(r)}{r}}, 
                              \scale[1]{\frac{U_{\rm LOS}k_{\xi}}{\sqrt{1+r^2}}}, 
                              \scale[1]{\frac{U_{\rm LOS}}{\sqrt{\Delta_0^2+2r^2}}}\right\}
        \end{equation}
        All assumptions of \cite[Theorem 5]{pettersen_lyapunov_2017} are thus satisfied, and the origin of the nominal system is USGES.

        \vspace{-0.2em}

        Moreover, note that the low-level controller is GES by Assumption \ref{ass5}.
        Therefore, if the following two assumptions hold, the origin of the cascade is USGES \citep[Proposition 9]{pettersen_lyapunov_2017}:
        \vspace{-0.3em}
        \begin{enumerate}
            \item There exist three positive constants $c_1, c_2, \eta$ such that
            \begin{align}
                \norm{\frac{\partial V}{\partial \widetilde{\bs{\sigma}}}} \norm{\widetilde{\bs{\sigma}}} &\leq c_1 V(\bs{\sigma}_1), & 
                \forall \norm{\widetilde{\bs{\sigma}}} &\geq \eta, \\
                \norm{\frac{\partial V}{\partial \widetilde{\bs{\sigma}}}} &\leq c_2, &
                \forall \norm{\widetilde{\bs{\sigma}}} &\leq \eta,
            \end{align}
            \item There exist two continuous functions $\alpha_{1}, \alpha_{2} : \mathbb{R}_{\geq 0} \mapsto \mathbb{R}_{\geq 0}$ such that
        \end{enumerate}
        \vspace{-0.5em}
        \begin{equation}
            \norm{\inlinevector{\mat{J}_2\mat{G}, \frac{1}{n}\sum_{i=1}^n\mat{g}_i}} \leq \alpha_1\left(\bigl\|\widetilde{\mat{X}}\bigr\|\right) + \alpha_2\left(\bigl\|\widetilde{\mat{X}}\bigr\|\right) \norm{\widetilde{\bs{\sigma}}}.
            \label{eq:USGES_assumption_2}
        \end{equation}
        Since $\norm{\partial V / \partial \widetilde{\bs{\sigma}}} = \norm{\widetilde{\bs{\sigma}}}$, the first assumption is satisfied for $c_1 = \sfrac{1}{2}$, $c_2 = \eta$, and any $\eta \in \mathbb{R}_{\geq 0}$.

        \vspace{-0.2em}

        To validate the second assumption, we first need to investigate the perturbing terms $\mat{g}_i$ from \eqref{eq:g_i}.
        From \eqref{eq:v_NSB_i} we get the following upper bound on $\bs{\upsilon}_{{\rm NSB}, i}$ \vspace{-0.15em}
        \begin{align}
            \norm{\bs{\upsilon}_{{\rm NSB}, i}} &\leq U_{\rm LOS}\left(1 + k_{\rm NSB}\right) + \upsilon_{2, \max}\tanh\left(\norm{\widetilde{\bs{\sigma}}_2}\right),
        \end{align}
        and from \eqref{eq:expm}, we get the inequalities \vspace{-0.15em}
        \begin{align}
            s &\leq 1, &
            \norm{c\,\bs{\delta}} &\leq \sfrac{\sqrt{2}}{2}.
        \end{align}
        Therefore, $\mat{g}_i$ can be upper-bounded by \vspace{-0.15em}
        \begin{equation}
            \norm{\mat{g}_i} \leq \norm{\bs{\upsilon}_{{\rm NSB}, i}}\left(1 + \sfrac{\sqrt{2}}{2}\right)\norm{\bs{\delta}_i} + \abs{\widetilde{u}_i}.
        \end{equation}
        Consider then the two functions $\alpha_{1,i}, \alpha_{2,i} : \mathbb{R}_{\geq 0} \mapsto \mathbb{R}_{\geq 0}$ \vspace{-0.15em}
        \begin{align}
            \alpha_{1,i}(r) &= \left(U_{\rm LOS}\left(1 + k_{\rm NSB}\right)\left(1 + \sfrac{\sqrt{2}}{2}\right) + 1\right)\,r, \\
            \alpha_{2,i}(r) &= \upsilon_{2, \max}\left(1 + \sfrac{\sqrt{2}}{2}\right)\,r.
        \end{align}
        Then, the following holds: \vspace{-0.45em}
        \begin{equation}
            \norm{\mat{g}_i} \leq \alpha_{1,i}\left(\bigl\|\widetilde{\mat{X}}_i\bigr\|\right) + 
                                \alpha_{2,i}\left(\bigl\|\widetilde{\mat{X}}_i\bigr\|\right)\norm{\widetilde{\bs{\sigma}}}. \label{eq:perturbing_term}
        \end{equation}
        Therefore, \eqref{eq:USGES_assumption_2} can be satisfied by \vspace{-0.25em}
        \begin{align}
            \alpha_1(r) &= \sum_{i=1}^n \alpha_{1,i}(r), &
            \alpha_2(r) &= \sum_{i=1}^n \alpha_{2,i}(r),
        \end{align}
        and consequently all assumptions of \cite[Proposition~9]{pettersen_lyapunov_2017} are satisfied. To summarize, the origin of the closed-loop system is USGES.

        \vspace{-0.35em}

        As for the underactuated dynamics, the assumption $\widetilde{\mat{X}} = \mat{0}$ implies $\bs{\omega}_i = \bs{\omega}_{d, i}$ and $u_i = u_{d, i}$.
        Therefore the underactuated dynamics depend on the desired angular velocity.
        Recall the definition of $\bs{\omega}_{d, i}$ in~\eqref{eq:omega_ref}.
        To find a closed-loop expression for $\bs{\omega}_{d, i}$, we shall analyze $\bs{\omega}_{\bs{\upsilon}_{{\rm NSB}, i}}$ and $\bs{\omega}_{\mat{v}_{i}}$.

        \vspace{-0.15em}
        
        First, we consider $\bs{\omega}_{\bs{\upsilon}_{{\rm NSB}, i}}$.
        In Appendix~\ref{app:v_NSB}, we show that there exist positive constants $a_{\rm NSB}$ and $b_{\rm NSB}$ such that \vspace{-0.2em}
        \begin{equation}
            \norm{\bs{\omega}_{\bs{\upsilon}_{{\rm NSB}, i}}} \leq a_{\rm NSB}\norm{\mat{v}_u} + b_{\rm NSB}.
            \label{eq:omega_NSB_bound}
        \end{equation}
        Now, let us consider $\bs{\omega}_{\mat{v}_i}$.
        In Appendix~\ref{app:omega_v}, we show that $\bs{\omega}_{\mat{v}_i}$ depends on the angular velocities of the vehicle, thus forming an algebraic loop.
        However, under certain conditions, this loop can be resolved.

        \vspace{-0.25em}

        We show that $\bs{\omega}_{\mat{v}_{i}}$ is affine in $\bs{\omega}_i$.
        In other words, there exist $\bs{\omega}_{0, i}$ and $\mat{A}_{\bs{\omega}_i}$ such that \vspace*{-0.35em}
        \begin{equation}
            \bs{\omega}_{\mat{v}_{i}} = \bs{\omega}_{0, i} + \mat{A}_{\bs{\omega}_i}\,\bs{\omega}_i.
            \vspace*{-0.15em}
        \end{equation}
        Moreover, we show that $\mat{A}_{\bs{\omega}_i}$ satisfies \vspace{-0.25em}
        \begin{equation}
            \det\left(\mat{I} + \mat{A}_{\bs{\omega}_i}\right) \geq 1 - k_a,
            \label{eq:det_A_bound}
            \vspace*{-0.25em}
        \end{equation}
        where $k_a$ is a positive constant depending on the physical properties of the vehicle, the minimum surge velocity, and the ocean current.
        If $k_a < 1$, then $\left(\mat{I} + \mat{A}_{\bs{\omega}_i}\right)$ is invertible, and the desired angular velocity is \vspace{-0.45em}
        \begin{equation}
            \bs{\omega}_{d, i} = \left(\mat{I} + \mat{A}_{\bs{\omega}_i}\right)^{-1}\left(\mat{R}_{d, i}\T \bs{\omega}_{\bs{\upsilon}_{{\rm NSB}, i}} - \bs{\omega}_{0, i}\right).
        \end{equation}
        In addition, there exist positive constants $a_v$, and $b_v$ such that \vspace{-0.15em}
        \begin{equation}
            \norm{\bs{\omega}_{0, i}} \leq a_v\norm{\mat{v}_u} + b_v.
            \label{eq:omega_0_bound}
        \end{equation}

        By combining \eqref{eq:omega_NSB_bound}, \eqref{eq:det_A_bound}, and \eqref{eq:omega_0_bound}, we can upper bound the angular velocity with \vspace{-0.15em}
        \begin{equation}
            \norm{\bs{\omega}_{d, i}} \leq \frac{\left(a_{\rm NSB} + a_v\right)\norm{\mat{v}_u} + b_{\rm NSB} + b_v}{1 - k_a} .
        \end{equation}
        The Lyapunov function candidate \vspace{-0.25em}
        \begin{equation}
            V_u = \frac{1}{2} \mat{v}_u\T\mat{v}_u
        \end{equation}
        for the underactuated dynamics may then be shown that, leveraging \eqref{eq:underactuated_dynamics_closed_loop}, has its time-derivative bounded by \vspace{-0.25em}
        \begin{equation}
            \dot{V}_u \leq \mat{v}_u\T\mat{Y}\mat{v}_u + aX_{\max}\norm{\mat{v}_u}^2 + H\left(\norm{\mat{v}_u}, \norm{\mat{V}_c}\right),
        \end{equation}
        where $a = (a_{\rm NSB} + a_v)/(1 - k_a)$, $X_{\max}$ is the largest singular value of $\mat{X}$, and $H$ represents the terms that grow at most linearly with $\mat{v}_u$.
        Since $\mat{Y}$ contains terms associated with hydrodynamic damping, it is negative definite.
        Therefore, $\dot{V}_u$ can be further bounded by \vspace{-0.25em}
        \begin{equation}
            \dot{V}_u \leq -\left(Y_{\min} - aX_{\max}\right)\norm{\mat{v}_u}^2 + H(\cdot),
        \end{equation}
        where $Y_{\min}$ is the real part of the smallest eigenvalue of $-\mat{Y}$.
        For a sufficiently large $\mat{v}_u$, the quadratic terms will dominate the linear terms.
        Consequently, the underactuated dynamics are bounded if \vspace{-0.15em}
        \begin{equation}
            Y_{\min} > aX_{\max}. \label{eq:boudedness_condition}
            \qed
        \end{equation}
        \vspace{-2em}
    \end{pf}

    \section{Simulations}
    \label{sec:simulation}
    \vspace{-1.1em}
    \begin{figure*}[ht]
        \centering
        \begin{subfigure}[t]{0.46\textwidth}
            \centering
            % This file was created by matlab2tikz.
%
%The latest updates can be retrieved from
%  http://www.mathworks.com/matlabcentral/fileexchange/22022-matlab2tikz-matlab2tikz
%where you can also make suggestions and rate matlab2tikz.
%
\definecolor{mycolor1}{RGB}{27,158,119}%
\definecolor{mycolor2}{RGB}{233,95,2}%
\definecolor{mycolor3}{RGB}{117,111,179}%
\definecolor{mycolor4}{rgb}{0.50000,1.00000,1.00000}%
\begin{tikzpicture}

\begin{axis}[%
width=65mm,
height=30mm,
at={(0mm, 0mm)},
scale only axis,
xmin=0,
xmax=250,
xlabel style={font=\color{white!15!black}, yshift=1mm},
xlabel={Time [s]},
xtick={0,50,100,150,200,250},
ymin=0,
ymax=25,
ylabel style={font=\color{white!15!black}},
ylabel={Distance [m]},
axis background/.style={fill=white},
title style={font=\bfseries, yshift=-2.5mm},
title={Smallest distances},
legend style={at={(0.97,0.05)}, anchor=south east, legend cell align=left, align=left, draw=white!15!black, legend columns=2}
]
\addplot [color=mycolor1, line width=1pt]
  table[]{simout-1.tsv};
\addlegendentry{Inter-vehicle}

\addplot [color=mycolor2, line width=1pt]
  table[]{simout-2.tsv};
\addlegendentry{Obstacle}

\addplot [color=black, dashed, line width=1.2pt]
  table[]{simout-3.tsv};
\addlegendentry{$d_{\rm COLAV} = r_o$}

\end{axis}
\end{tikzpicture}%
            \vspace{-3mm}
            \caption{The smallest inter-vehicle and vehicle-to-obstacle distance.}
            \label{fig:distances}
            \vspace{-1mm}
        \end{subfigure}
        \hspace{1em}
        \begin{subfigure}[t]{0.46\textwidth}
            \centering
            % This file was created by matlab2tikz.
%
%The latest updates can be retrieved from
%  http://www.mathworks.com/matlabcentral/fileexchange/22022-matlab2tikz-matlab2tikz
%where you can also make suggestions and rate matlab2tikz.
%
\definecolor{mycolor1}{RGB}{27,158,119}%
\definecolor{mycolor2}{RGB}{233,95,2}%
\definecolor{mycolor3}{RGB}{117,111,179}%
\definecolor{mycolor4}{rgb}{0.50000,1.00000,1.00000}%
\begin{tikzpicture}

\begin{axis}[%
width=65mm,
height=30mm,
at={(0mm, 0mm)},
scale only axis,
xmin=0,
xmax=250,
xlabel style={font=\color{white!15!black}, yshift=1mm},
xlabel={Time [s]},
xtick={0,50,100,150,200,250},
y dir=reverse,
ymin=0,
ymax=60,
ylabel style={font=\color{white!15!black}},
ylabel={$z$-coordinate [m]},
axis background/.style={fill=white},
title style={font=\bfseries, yshift=-2.5mm},
title={Depth},
legend style={at={(0.03,0.95)}, anchor=north west, legend cell align=left, align=left, draw=white!15!black}
]
\addplot [color=mycolor1, line width=1pt]
    table[]{simout-4.tsv};
\addlegendentry{Smallest depth}

\addplot [color=mycolor2, line width=1pt]
    table[]{simout-5.tsv};
\addlegendentry{Largest depth}

\addplot [color=black, dashed, line width=1.2pt]
    table[]{simout-6.tsv};
\addlegendentry{Depth limits}

\addplot [color=black, dashed, forget plot, line width=1.2pt]
    table[]{simout-7.tsv};
\end{axis}
\end{tikzpicture}
            \vspace{-3mm}
            \caption{The smallest and largest vehicle depth.}
            \label{fig:depth}
            \vspace{-1mm}
        \end{subfigure}
        \begin{subfigure}[t]{0.46\textwidth}
            \centering
            % This file was created by matlab2tikz.
%
%The latest updates can be retrieved from
%  http://www.mathworks.com/matlabcentral/fileexchange/22022-matlab2tikz-matlab2tikz
%where you can also make suggestions and rate matlab2tikz.
%
\definecolor{mycolor1}{RGB}{27,158,119}%
\definecolor{mycolor2}{RGB}{233,95,2}%
\definecolor{mycolor3}{RGB}{117,111,179}%
\definecolor{mycolor4}{rgb}{0.50000,1.00000,1.00000}%
\begin{tikzpicture}

\begin{axis}[%
width=65mm,
height=25mm,
at={(0mm, 0mm)},
scale only axis,
xmin=0,
xmax=250,
xlabel style={font=\color{white!15!black}, yshift=1mm},
xtick={0,50,100,150,200,250},
xlabel={Time [s]},
ymin=-20.4602266733107,
ymax=24.5,
ylabel style={font=\color{white!15!black}, yshift=-2mm},
ylabel={Error [m]},
axis background/.style={fill=white},
title style={font=\bfseries, yshift=-2.5mm},
title={Path following error},
legend style={at={(0.95,1.)}, anchor=north east, legend cell align=left, align=left, draw=white!15!black}
]
\addplot [color=mycolor1, line width=1pt]
    table[]{simout-8.tsv};
\addlegendentry{$x$-error}

\addplot [color=mycolor2, line width=1pt]
    table[]{simout-9.tsv};
\addlegendentry{$y$-error}

\addplot [color=mycolor3, line width=1pt]
    table[]{simout-10.tsv};
\addlegendentry{$z$-error}

\addplot[area legend, dashed, draw=black, fill=green, fill opacity=0.15, forget plot]
table[] {simout-11.tsv}--cycle;

\addplot[area legend, dashed, draw=black, fill=orange, fill opacity=0.15, forget plot]
table[] {simout-12.tsv}--cycle;

\addplot[area legend, dashed, draw=black, fill=orange, fill opacity=0.15, forget plot]
table[] {simout-13.tsv}--cycle;
\end{axis}
\end{tikzpicture}
            \vspace{-3mm}
            \caption{The path-following error. The green rectangle represents the time when obstacle avoidance is active. The red rectangle represents the time when depth limiting is active.}
            \label{fig:path_following_error}
            \vspace{-1mm}
        \end{subfigure}
        \hspace{1em}
        \begin{subfigure}[t]{0.46\textwidth}
            \centering
            % This file was created by matlab2tikz.
%
%The latest updates can be retrieved from
%  http://www.mathworks.com/matlabcentral/fileexchange/22022-matlab2tikz-matlab2tikz
%where you can also make suggestions and rate matlab2tikz.
%
\definecolor{mycolor1}{RGB}{27,158,119}%
\definecolor{mycolor2}{RGB}{233,95,2}%
\definecolor{mycolor3}{RGB}{117,111,179}%
\definecolor{mycolor4}{rgb}{0.50000,1.00000,1.00000}%
\begin{tikzpicture}

\begin{axis}[%
width=65mm,
height=25mm,
at={(0mm, 0mm)},
scale only axis,
xmin=0,
xmax=250,
xlabel style={font=\color{white!15!black}, yshift=1mm},
xlabel={Time [s]},
xtick={0,50,100,150,200,250},
ymin=-25,
ymax=25,
ylabel style={font=\color{white!15!black}, yshift=-2mm},
ylabel={Error [m]},
axis background/.style={fill=white},
title style={font=\bfseries, yshift=-2.5mm},
title={Formation keeping error},
legend style={at={(0.97,0.03)}, anchor=south east, legend cell align=left, align=left, draw=white!15!black, legend columns=1}
]
\addplot [color=mycolor1, line width=1pt]
    table[]{simout-14.tsv};
    \addlegendentry{$x$-error}
\addplot [color=mycolor1, dashed, forget plot, line width=1.1pt]
    table[]{simout-15.tsv};
\addplot [color=mycolor1, dotted, forget plot, line width=1.3pt]
    table[]{simout-16.tsv};
\addplot [color=mycolor2, line width=1pt]
    table[]{simout-17.tsv};
    \addlegendentry{$y$-error}
\addplot [color=mycolor2, dashed, forget plot, line width=1.1pt]
    table[]{simout-18.tsv};
\addplot [color=mycolor2, dotted, forget plot, line width=1.3pt]
    table[]{simout-19.tsv};
\addplot [color=mycolor3, line width=1pt]
    table[]{simout-20.tsv};
    \addlegendentry{$z$-error}
\addplot [color=mycolor3, dashed, forget plot, line width=1.1pt]
    table[]{simout-21.tsv};
\addplot [color=mycolor3, dotted, forget plot, line width=1.3pt]
    table[]{simout-22.tsv};

\addplot[area legend, dashed, draw=black, fill=mycolor4, fill opacity=0.25, forget plot]
table[] {simout-23.tsv}--cycle;
\end{axis}

\end{tikzpicture}
            \vspace{-3mm}
            \caption{The formation-keeping error. The full, dashed, and dotted lines correspond to vehicles 1, 2, and 3, respectively. The blue rectangle represents the time when inter-agent COLAV is active.}
            \label{fig:formation_keeping_error}
            \vspace{-1mm}
        \end{subfigure}
        \begin{subfigure}[t]{0.46\textwidth}
            \centering
            % This file was created by matlab2tikz.
%
%The latest updates can be retrieved from
%  http://www.mathworks.com/matlabcentral/fileexchange/22022-matlab2tikz-matlab2tikz
%where you can also make suggestions and rate matlab2tikz.
%
\definecolor{mycolor1}{RGB}{27,158,119}%
\definecolor{mycolor2}{RGB}{233,95,2}%
\definecolor{mycolor3}{RGB}{117,111,179}%
\definecolor{mycolor4}{rgb}{0.50000,1.00000,1.00000}%
\begin{tikzpicture}

\begin{axis}[%
width=65mm,
height=25mm,
at={(0mm, 0mm)},
scale only axis,
xmin=0,
xmax=250,
xlabel style={font=\color{white!15!black}, yshift=1mm},
xlabel={Time [s]},
xtick={0,50,100,150,200,250},
ymin=0,
ymax=1.4,
ylabel style={font=\color{white!15!black}, yshift=-2mm},
ylabel={Velocity [m/s]},
axis background/.style={fill=white},
title style={font=\bfseries, yshift=-2.5mm},
title={Surge velocity},
legend style={at={(0.97,0.07)}, anchor=south east, legend cell align=left, align=left, draw=white!15!black, legend columns=4}
]
\addplot [color=mycolor1, line width=1pt]
  table[]{states-1.tsv};
  \addlegendentry{$u_1$}
\addplot [color=mycolor2, line width=1pt]
  table[]{states-2.tsv};
  \addlegendentry{$u_3$}
\addplot [color=mycolor3, line width=1pt]
  table[]{states-3.tsv};
  \addlegendentry{$u_2$}
\addplot [color=black, dashed, line width=1.2pt]
  table[]{states-10.tsv};
  \addlegendentry{$u_{\min}$}

\addplot[area legend, dotted, draw=black, fill=gray, fill opacity=0.15, forget plot]
table[] {states-11.tsv}--cycle;

\addplot[area legend, dotted, draw=black, fill=gray, fill opacity=0.15, forget plot]
table[] {states-12.tsv}--cycle;
\end{axis}
\end{tikzpicture}
            \vspace{-3mm}
            \caption{The surge velocities of the vehicles. The grey rectangle represents the time when any avoidance task is active.}
            \label{fig:surge}
            \vspace{-2.5mm}
        \end{subfigure}
        \hspace{1em}
        \begin{subfigure}[t]{0.46\textwidth}
            \centering
            % This file was created by matlab2tikz.
%
%The latest updates can be retrieved from
%  http://www.mathworks.com/matlabcentral/fileexchange/22022-matlab2tikz-matlab2tikz
%where you can also make suggestions and rate matlab2tikz.
%
\definecolor{mycolor1}{RGB}{27,158,119}%
\definecolor{mycolor2}{RGB}{233,95,2}%
\definecolor{mycolor3}{RGB}{117,111,179}%
\definecolor{mycolor4}{rgb}{0.50000,1.00000,1.00000}%
\begin{tikzpicture}

\begin{axis}[%
width=65mm,
height=25mm,
at={(0mm, 0mm)},
scale only axis,
xmin=0,
xmax=250,
xlabel style={font=\color{white!15!black}, yshift=1mm},
xlabel={Time [s]},
xtick={0,50,100,150,200,250},
ymin=-0.1,
ymax=0.2,
ylabel style={font=\color{white!15!black}, yshift=-4mm},
ylabel={Velocity [m/s]},
axis background/.style={fill=white},
title style={font=\bfseries, yshift=-2.5mm},
title={Underactuated dynamics},
legend style={at={(0.97,0.05)}, anchor=south east, legend cell align=left, align=left, draw=white!15!black, legend columns=3}
]
\addplot [color=mycolor1, line width=1pt]
  table[]{states-4.tsv};
  \addlegendentry{$v$}
\addplot [color=mycolor1, dashed, forget plot, line width=1.1pt]
  table[]{states-5.tsv};  
\addplot [color=mycolor1, dotted, forget plot, line width=1.3pt]
  table[]{states-6.tsv}; 

\addplot [color=mycolor2, line width=1pt]
  table[]{states-7.tsv};
  \addlegendentry{$w$}
\addplot [color=mycolor2, dashed, forget plot, line width=1.1pt]
  table[]{states-8.tsv};  
\addplot [color=mycolor2, dotted, forget plot, line width=1.3pt]
  table[]{states-9.tsv}; 

\addplot[area legend, dotted, draw=black, fill=gray, fill opacity=0.15, forget plot]
  table[] {states-22.tsv}--cycle;  
\addplot[area legend, dotted, draw=black, fill=gray, fill opacity=0.15, forget plot]
  table[] {states-23.tsv}--cycle;
\end{axis}

\end{tikzpicture}
            \vspace{-3mm}
            \caption{The sway and heave velocities. The full, dashed, and dotted lines correspond to vehicles 1, 2, and 3, respectively.}
            \label{fig:sway_heave}
            \vspace{-2.5mm}
        \end{subfigure}
        \caption{Simulation results.}
        \label{fig:results}
        \vspace{-0.7em}
    \end{figure*}
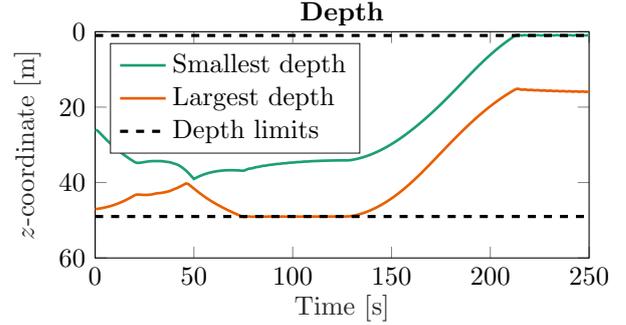
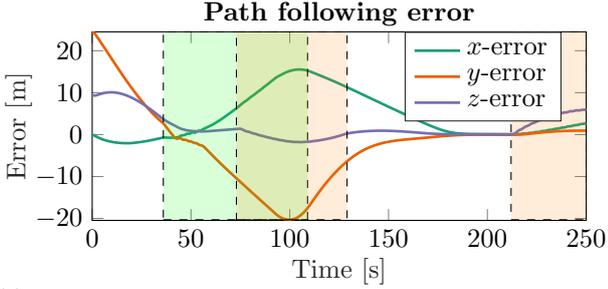
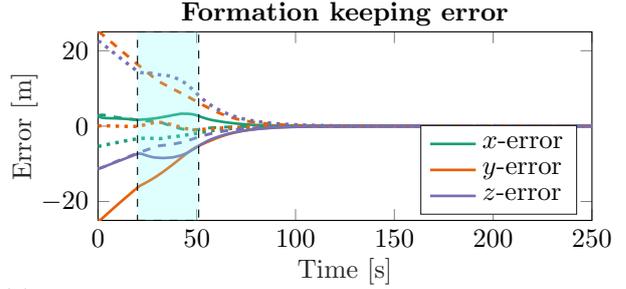
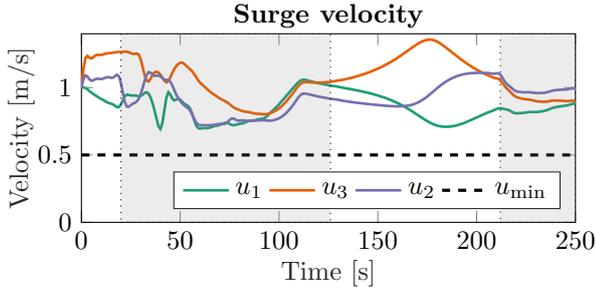
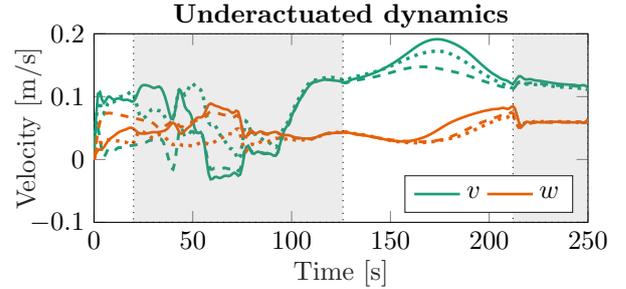

    \begin{figure}[b]
        \centering
        \includegraphics[width=.45\textwidth]{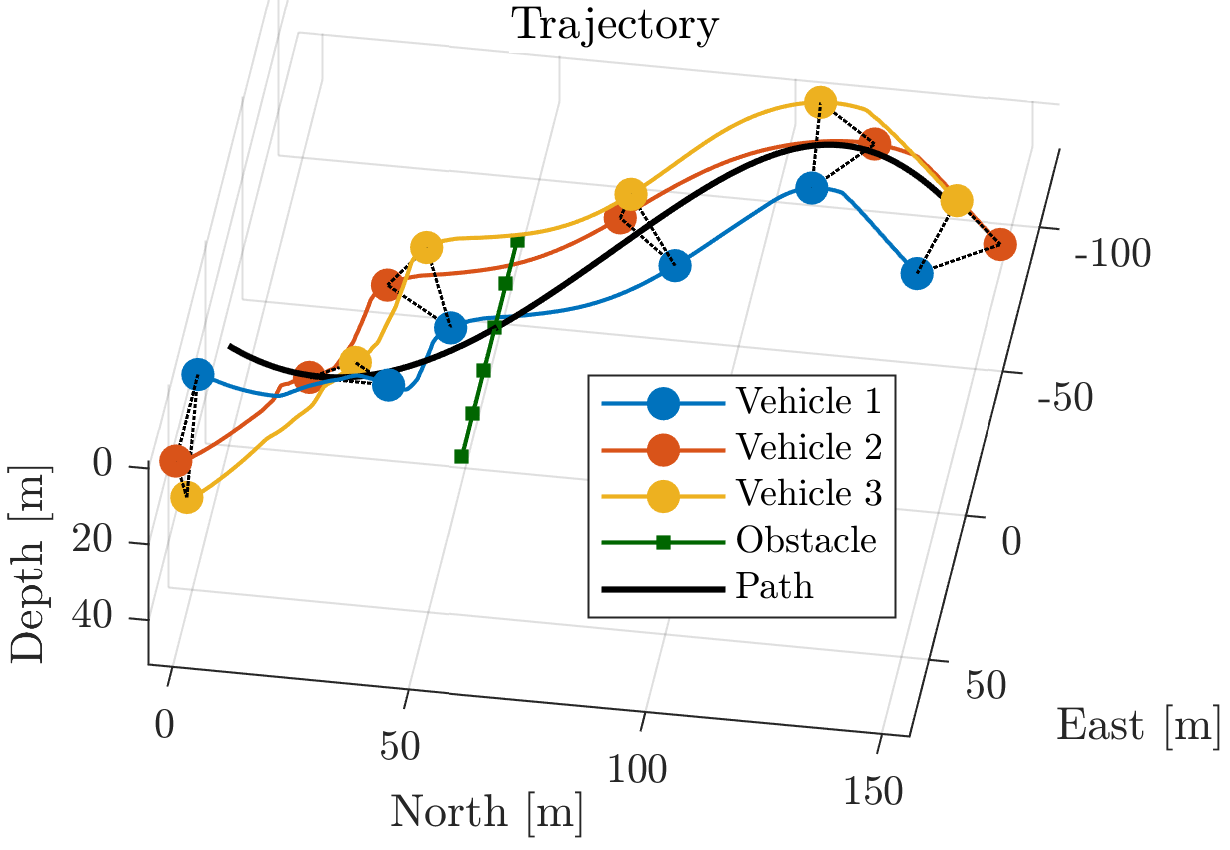}
        \vspace{-4mm}
        \caption{The 3D trajectory of the vehicles. The markers represent the position of the vehicles at times $t = 0, 50, \ldots, 250$ seconds. Markers with corresponding times are connected by dotted lines to better illustrate the resulting formation.}
        \label{fig:trajectory}
    \end{figure}

    We simulate the proposed approach on a fleet of six LAUVs \citep{sousa_LAUV_2012} using MATLAB, delegating low-level control to an attitude-tracking PID controller as in \cite{nakath_rigid_2017} and an output-linearizing P surge controller as in \cite{matouvs_formation_2022}.

    \vspace{-0.35em}

    The desired path is a spiral given by \vspace{-0.15em}
    \begin{equation}
        \mat{p}_p(\xi) = \mat{p}_{p, 0} + \inlinevector{\xi, a_p\,\cos(\omega_p\,\xi), b_p\,\sin(\omega_p\,\xi)}
    \end{equation}
    where \vspace{-0.45em}
    \begin{align*}
        \mat{p}_{p, 0} &= \inlinevector{0, -40, 25}, &
        a_p &= 40, &
        b_p &= 20, &
        \omega_p &= \scale[1]{\frac{100}{\pi}},
    \end{align*}
    while the desired formation is an isosceles triangle parallel to the $yz$ plane.
    Specifically, the desired positions in the formation-centered frame are \vspace{-0.15em}
    \begin{align}
        \mathbf{p}_{f,1}^f &= \begin{bmatrix} 0 \\ 10 \\ 5\end{bmatrix}, &
        \mathbf{p}_{f,2}^f &= \begin{bmatrix} 0 \\ -10 \\ 5\end{bmatrix}, &
        \mathbf{p}_{f,3}^f &= \begin{bmatrix} 0 \\ 0 \\ -10\end{bmatrix}.
    \end{align}
    For the simulation parameters, we choose the velocity of the ocean current to be $\mat{V}_c = \inlinevector{0, 0.15, 0.05}$, the formation-keeping gain $\bs{\Lambda}_2 = 0.1\mat{I}$, the maximum formation-keeping velocity $\upsilon_{2, \max} = \SI{0.5}{\meter\per\second}$, and the lookahead distance $\Delta_0 = \SI{5}{\meter}$.

    \vspace{-0.35em}
    
    The very minimum relative distance to avoid collision is the length of the LAUV, \emph{i.e.}, $2.4$ m.
    For additional safety, we design the COLAV task with $d_{\rm min} = 5$ m.
    For additional safety during transients, $d_{\rm COLAV}$ is chosen to be $10$ m.

    \vspace{-0.35em}

    We then let the vehicles encounter an obstacle of similar size as the LAUV that moves east at a constant speed of $\SI{0.3}{\meter\per\second}$.
    Given its size, we choose $r_o = d_{\rm COLAV}$.
    The minimum cone angle is set to $\alpha_{\min} = \SI{15}{\degree}$.
    The operation limits are chosen as $z_{\min} = \SI{1}{m}$, $z_{\max} = \SI{49}{m}$, and the depth-limiting velocity is $\upsilon_z = \SI{0.3}{\meter\per\second}$.
    Note that the limits are deliberately chosen too small for the given path and formation, so that depth limiting is activated.

    \vspace{-0.3em}

    Figures \ref{fig:results} and \ref{fig:trajectory} show the results of this numerical simulation.
    %The initial conditions are chosen identically as in \cite{matouvs_formation_2022}, so that we can directly compare the results.
    \reffig{fig:distances} shows the distance between the vehicles and the distance to the obstacle.
    At $t = \SI{20}{\second}$, the COLAV task is activated, and the distance between the vehicles drops to approximately $9.5$ meters during the transient.
    The situation is resolved after $30$ seconds.
    At $t = \SI{35}{\second}$, the vehicles enter the collision cone and perform an evasive maneuver in a clockwise direction.
    The distance to the obstacle is always above the required limit.

    \vspace{-0.3em}

    \reffig{fig:depth} shows the depth of the vehicles.
    At $t = \SI{73}{\second}$ and $t = \SI{212}{\second}$, the depth-limiting task is activated.
    When the task is active, the depth of the vehicles fluctuates around the prescribed limit.

    \vspace{-0.3em}

    Figures \ref{fig:path_following_error} and \ref{fig:formation_keeping_error} show the path-following and formation-keeping errors.
    We can see that the path-following errors diverge when obstacle avoidance or depth limiting is active.
    Conversely, the formation-keeping errors diverge during inter-agent COLAV.
    This behavior corresponds to the interpretation of the NSB tasks --- path-following is global and thus cannot be satisfied during obstacle avoidance, whereas formation-keeping works with relative velocities and thus cannot be satisfied during inter-agent COLAV.

    \vspace{-0.3em}

    \reffig{fig:surge} shows the surge velocity of the vehicles.
    We can see that the surge velocities are always above the required limit.
    In fact, our solution appears to be overly conservative.
    \reffig{fig:sway_heave} shows the sway and heave velocities.
    We can see that the velocities change abruptly when the collision avoidance or depth limiting tasks are active, as the vehicles switch to a different behavior.
    However, the velocities still remain bounded during the whole simulation.
    The peak in sway velocities at $t = \SI{180}{\second}$ coincides with the sharpest turn (\emph{i.e.,} the largest $\bs{\omega}_p(\xi)$) of the desired path.

    \vspace{-0.8em}
    \section{Conclusions and Future Work}
    \label{sec:conclusion}
    \vspace{-1.2em}
    This paper extends a formation path-following NSB algorithm to underactuated 6DOF vehicles while adding obstacle avoidance and depth-limiting capabilities. 
    Both the path-following and formation-keeping parts are proven to be stable.     
    In the proofs, we assume that the avoidance and depth-limiting tasks are not active.
    An analysis of the closed-loop system with active avoidance and depth-limiting tasks is left for future work.
    \vspace{-0.8em}
    
    \bibliography{biblio}

    \onecolumn
    \appendix
    \section{}
    \label{appendix}

    \subsection{Bounds on $\bs{\omega}_{\bs{\upsilon}_{{\rm NSB}, i}}$}
    \label{app:v_NSB}
    Recall the definition of $\bs{\omega}_{\bs{\upsilon}_{{\rm NSB},i}}$ in \eqref{eq:omega_v}.
    Note that by definition, a normalized vector is always orthogonal to its derivative.
    Therefore, the following equality holds:
    \begin{equation}
        \norm{\bs{\omega}_{\bs{\upsilon}_{{\rm NSB}, i}}} = \norm{\overline{\bs{\upsilon}}_{{\rm NSB}, i}}\norm{\dot{\overline{\bs{\upsilon}}}_{{\rm NSB}, i}}
        = \norm{\dot{\overline{\bs{\upsilon}}}_{{\rm NSB}, i}}.
    \end{equation}
    Therefore, instead of the pseudo-angular velocity, it is possible to investigate the derivative of the normalized NSB velocity.
    Note that according to the assumptions in Theorem~\ref{thm1}, the analysis should be performed on the manifold $\inlinevector{\tilde{\bs{\sigma}}, \tilde{\mat{X}}} = \mat{0}$.
    Substituting $\tilde{\bs{\sigma}} = \mat{0}$ to \eqref{eq:v_NSB_i} yields
    \begin{equation}
        \bs{\upsilon}_{{\rm NSB}, i} = \bs{\upsilon}_{\rm LOS} + \dot{\mat{R}}_p(\xi)\mat{p}_{f,i}^f = 
        U_{\rm LOS}\mat{R}_p(\xi) \left(\mat{e}_1 + \norm{\partial \mat{p}_p(\xi) / \partial \xi}^{-1}\bs{\omega}_p(\xi) \times \mat{p}_{f,i}^f\right).
        \label{eq:v_NSB_i_nominal}
    \end{equation}
    For brevity, let us define
    \begin{align}
        \bs{\kappa} &= \norm{\partial \mat{p}_p(\xi) / \partial \xi}^{-1}\bs{\omega}_p(\xi), &
        \mat{e}_p &= \mat{e}_1 + \bs{\kappa} \times \mat{p}_{f,i}^f
    \end{align}
    The normalized NSB velocity is then given by
    \begin{equation}
        \overline{\bs{\upsilon}}_{{\rm NSB}, i} = \frac{\mat{R}_p(\xi) \mat{e}_p}{\norm{\mat{e}_p}}.
        \label{eq:v_NSB_i_normalized}
    \end{equation}
    Differentiating \eqref{eq:v_NSB_i_normalized} with respect to time yields
    \begin{equation}
        \dot{\overline{\bs{\upsilon}}}_{{\rm NSB}, i} = 
        \frac{U_{\rm LOS}\mat{R}_p\left(\bs{\kappa}\times\mat{e}_p + \bs{\iota}\times\mat{p}_{f,i}^f\right)}{\norm{\mat{e}_p}}
        - \frac{U_{\rm LOS}\mat{R}_p\mat{e}_p\left(\mat{e}_p\T\left(\bs{\iota}\times\mat{p}_{f,i}^f\right)\right)}{\norm{\mat{e}_p}^2},
        \label{eq:v_NSB_i_dot}
    \end{equation}
    where $\bs{\iota} = \partial \bs{\kappa} / \partial \xi$. From \eqref{eq:v_NSB_i_dot}, it follows that
    \begin{equation}
        \norm{\dot{\overline{\bs{\upsilon}}}_{{\rm NSB}, i}} \leq
        U_{\rm LOS} \left(\norm{\bs{\kappa}} + \frac{\norm{\bs{\iota}\times\mat{p}_{f,i}^f}\left(1 + \norm{\mat{e}_p}\right)}{\norm{\mat{e}_p}}\right).
        \label{eq:v_NSB_i_dot_bound1}
    \end{equation}
    If we assume that the second and third partial derivatives of $\mat{p}_p$ with respect to the path parameter are bounded, then $\bs{\iota}$ is bounded as well.
    Let us define
    \begin{equation}
        c_{\rm NSB} = \max_{i, \xi} \left(\norm{\bs{\kappa}} + \frac{\norm{\bs{\iota}\times\mat{p}_{f,i}^f}\left(1 + \norm{\mat{e}_p}\right)}{\norm{\mat{e}_p}}\right).
        \label{eq:c_NSB}
    \end{equation}
    Substituting \eqref{eq:U_LOS} and \eqref{eq:c_NSB} into \eqref{eq:v_NSB_i_dot_bound1} gives us the following upper bound
    \begin{equation}
        \norm{\dot{\overline{\bs{\upsilon}}}_{{\rm NSB}, i}} \leq 
        \frac{\upsilon_{2, \max} + \sqrt{\sum_{i=1}^n \left(v_i^2 + w_i^2\right) + u_{\min}^2}}{1 - k_{\rm NSB}}\,c_{\rm NSB}.
        \label{eq:v_NSB_i_dot_bound2}
    \end{equation}
    Note that for any two positive numbers $a$ and $b$, the following inequality holds: $\sqrt{a + b} \leq \sqrt{a} + \sqrt{b}$.
    Therefore, we can further upper-bound \eqref{eq:v_NSB_i_dot_bound2} with
    \begin{equation}
        \norm{\dot{\overline{\bs{\upsilon}}}_{{\rm NSB}, i}} \leq 
        \underbrace{\frac{c_{\rm NSB}}{1 - k_{\rm NSB}}}_{a_{\rm NSB}}\norm{\mat{v}_u} +
        \underbrace{\frac{\upsilon_{2, \max} + u_{\min}}{1 - k_{\rm NSB}}\,c_{\rm NSB}}_{b_{\rm NSB}}.
    \end{equation}
    We have thus shown that there exist positive constants $a_{\rm NSB}$ and $b_{\rm NSB}$ that satisfy \eqref{eq:omega_NSB_bound}.

    \subsection{Bounds on $\bs{\omega}_{\mat{v}_i}$}
    \label{app:omega_v}
    Note that by the assumptions of Theorem~\ref{thm1}, the surge velocity of the vehicle satisfies $u_i = u_{d, i}$, and the linear velocity vector $\mat{v}_i$ thus  satisfies
    \begin{align}
        \mat{v}_i &= \inlinevector{u_{d,i}, v_i, w_i} = \inlinevector{\sqrt{\norm{\bs{\upsilon}_{{\rm NSB}, i}}^2 - v_i^2 - w_i^2}, v_i, w_i}, &
        \norm{\mat{v}_i} &= \norm{\bs{\upsilon}_{{\rm NSB}, i}}.
        \label{eq:v_i_nominal}
    \end{align}
    The time-derivative of a normalized vector is given by
    \begin{equation}
        \dot{\overline{\mat{v}}}_i = \frac{\dot{\mat{v}}_i}{\norm{\mat{v}_i}} - \frac{\mat{v}_i\,\,\frac{\rm d}{{\rm d}t}\!\norm{\mat{v}_i}}{\norm{\mat{v}_i}^2},
    \end{equation}
    and the pseudo-angular velocity is thus given by
    \begin{equation}
        \bs{\omega}_{\mat{v}_i} = \overline{\mat{v}}_i \times \dot{\overline{\mat{v}}}_i
        = \frac{\mat{v}_i}{\norm{\mat{v}_i}} \times 
            \left(\frac{\dot{\mat{v}}_i}{\norm{\mat{v}_i}} - \frac{\mat{v}_i\,\,\frac{\rm d}{{\rm d}t}\!\norm{\mat{v}_i}}{\norm{\mat{v}_i}^2}\right)
        = \frac{\mat{v}_i \times \dot{\mat{v}}_i}{\norm{\mat{v}_i}^2}.
        \label{eq:omega_v_a1}
    \end{equation}
    Now, let us focus on $\dot{\mat{v}}_i$.
    Differentiating \eqref{eq:v_i_nominal} with respect to time yields
    \begin{equation}
        \dot{\mat{v}}_i = \begin{bmatrix}
            \frac{\bs{\upsilon}_{{\rm NSB}, i}\T\dot{\bs{\upsilon}}_{{\rm NSB}, i} - v_i\dot{v}_i - w_i\dot{w}_i}{u_i} \\
            \dot{v}_i \\
            \dot{w}_i
        \end{bmatrix}.
        \label{eq:v_i_dot}
    \end{equation}
    From \eqref{eq:underactuated_dynamics}, the underactuated dynamics are given by
    \begin{subequations}
        \begin{align}
            \dot{v}_i &= \left(X_{v0} + X_{v1}(u_i-u_c)\right)r_i + \left(Y_{v0} + Y_{v1}(u_i-u_c)\right)(v_i-v_c) + \left(Z_{v0} + Z_{v1}p_i\right)(w_i-w_c) + w_cp_i - u_cr_i, \\
            \dot{w}_i &= \left(X_{w0} + X_{w1}(u_i-u_c)\right)q_i + \left(Y_{w0} + Y_{w1}(u_i-u_c)\right)(w_i-w_c) + \left(Z_{w0} + Z_{w1}p_i\right)(v_i-v_c) + u_cq_i - v_cp_i,
        \end{align}
        \label{eq:underactuated_dynamics_expanded}
    \end{subequations}
    where
    \begin{subequations}
        \begin{align}
            X_v(u_r) &= X_{v0} + X_{v1}u_r, &
            Y_v(u_r) &= Y_{v0} + Y_{v1}u_r, &
            Z_v(p) &= Z_{v0} + Z_{v1}p, \\
            X_w(u_r) &= X_{w0} + X_{w1}u_r, &
            Y_w(u_r) &= Y_{w0} + Y_{w1}u_r, &
            Z_w(p) &= Z_{w0} + Z_{w1}p.
        \end{align}
    \end{subequations}
    Substituting \eqref{eq:underactuated_dynamics_expanded} into \eqref{eq:v_i_dot} yields
    \begin{equation}
    \begin{split}
        \dot{\mat{v}}_i =&
        \underbrace{\begin{bmatrix}
            \frac{w_{i}\,\left(v_{c}-Z_{w1}\,v_{r}\right)-v_{i}\,\left(w_{c}+Z_{v1}\,w_{r}\right)}{u_{i}} & 
            -\frac{w_{i}\,\left(X_{w0}+X_{w1}\,u_{r}+u_{c}\right)}{u_{i}} & 
            \frac{v_{i}\,\left(u_{c}-X_{v0}-X_{v1}\,u_{r}\right)}{u_{i}} \\ 
            w_{c}+Z_{v1}\,w_{r} & 0 & X_{v0}+X_{v1}\,u_{r}-u_{c} \\ 
            -v_{c}-Z_{w1}\,v_{r} & X_{w0}+X_{w1}\,u_{r}+u_{c} & 0
        \end{bmatrix}}_{\widehat{\mat{A}}_{\bs{\omega}_i}}
        \begin{bmatrix}
            p_i \\ q_i \\ r_i
        \end{bmatrix} \\
        & + \underbrace{\begin{bmatrix}
            \frac{\bs{\upsilon}_{{\rm NSB}, i}\T\dot{\bs{\upsilon}}_{{\rm NSB}, i} 
            - v_i\left(\left(Y_{v0} + Y_{v1}u_r\right)v_r + Z_{v0}w_r\right)
            - w_i\left(\left(Y_{w0} + Y_{w1}u_r\right)w_r + Z_{w0}v_r\right)}{u_i} \\
            \left(Y_{v0} + Y_{v1}u_r\right)v_r + Z_{v0}w_r \\
            \left(Y_{w0} + Y_{w1}u_r\right)w_r + Z_{w0}v_r
        \end{bmatrix}}_{\widehat{\bs{\omega}}_{0, i}}.
    \end{split}
    \label{eq:v_i_dot_matrix_form}
    \end{equation}
    Substituting \eqref{eq:v_i_dot_matrix_form} into \eqref{eq:omega_v_a1} yields
    \begin{equation}
        \bs{\omega}_{\mat{v}_i} = \frac{\mat{v}_i \times \left(\widehat{\mat{A}}_{\bs{\omega}_i}\bs{\omega}_i + \widehat{\bs{\omega}}_{0, i}\right)}{\norm{\mat{v}_i}^2}
        = \underbrace{\frac{\mat{S}\left(\mat{v}_i\right)\widehat{\mat{A}}_{\bs{\omega}_i}}{\norm{\mat{v}_i}^2}}_{\mat{A}_{\bs{\omega}_i}}\bs{\omega}_i
         + \underbrace{\frac{\mat{v}_i \times \widehat{\bs{\omega}}_{0, i}}{\norm{\mat{v}_i}^2}}_{\bs{\omega}_{0, i}}.
         \label{eq:omega_v_affine}
    \end{equation}
    We have thus shown that $\bs{\omega}_{\mat{v}_i}$ is affine in $\bs{\omega}_i$.

    Now we investigate the determinant of $(\mat{I} + \mat{A}_{\bs{\omega}_i})$.
    From the definition of $\mat{A}_{\bs{\omega}_i}$ in \eqref{eq:omega_v_affine}, we get the following expression
    \begin{equation}
    \begin{split}
        \det\left(\mat{I} + \mat{A}_{\bs{\omega}_i}\right) = \bigg( &
        u_i\left(u_i^2 + v_i^2 + w_i^2\right) - u_c\left(u_i^2 + v_i^2 + w_i^2\right) - \left(u_cu_i + v_cv_i + w_cw_i\right)\left(u_i - u_c\right) + X_{v0}\left(u_i^2 + v_i^2\right) \\
        & - X_{w0}\left(u_i^2 + w_i^2\right) + \left(X_{v1} - X_{w1}\right)u_i\left(u_i - u_c\right)^2 + \left(X_{v1} + Z_{w1}\right)v_i^2\left(u_i - u_c\right) \\
        & - \left(X_{w1} + Z_{v1}\right)w_i^2\left(u_i - u_c\right) - X_{v0}X_{w0}u_i - X_{v0}\left(u_iu_c + v_iv_c\right) + X_{w0}\left(u_iu_c + w_iw_c\right) \\
        & - X_{v0}\left(X_{w1}u_i^2 - Z_{w1}v_i^2\right) - X_{w0}\left(X_{v1}u_i^2 - Z_{v1}w_i^2\right) - X_{v1}X_{w1}u_i\left(u_i - u_c\right)^2 \\
        & - \left(X_{v1} + Z_{w1}\right)v_iv_c\left(u_i - u_c\right) + \left(X_{w1} + Z_{v1}\right)w_iw_c\left(u_i - u_c\right) + X_{v1}Z_{w1}v_i^2\left(u_i - u_c\right) \\
        & + X_{w1}Z_{v1}w_i^2\left(u_i - u_c\right) + X_{v0}\left(X_{w1}u_iu_c - Z_{w1}v_iv_c\right) + X_{w0}\left(X_{v1}u_iu_c - Z_{v1}w_iw_c\right) \\
        & - X_{v1}Z_{w1}v_iv_c\left(u_i - u_c\right) - X_{w1}Z_{v1}w_iw_c\left(u_i - u_c\right)\bigg) \frac{1}{u_i\left(u_i^2 + v_i^2 + w_i^2\right)}.
    \end{split}
    \end{equation}
    We need to find an upper bound on this expression.
    To do so, we will employ the following strategy:
    If possible, we will cancel the terms in the denominator with terms in the numerator.
    If the terms cannot be canceled, we will use the fact that $u_i \geq u_{\min}$, and put the following upper bound on the denominator
    \begin{equation}
        \frac{1}{u_i\left(u_i^2 + v_i^2 + w_i^2\right)} \leq \frac{1}{u_{\min}^3}.
    \end{equation}
    Furthermore, we will utilize the following inequalities that hold for any $a, b, c, K, L \in \mathbb{R}$
    \begin{subequations}
    \begin{align}
        \abs{a} &\leq \sqrt{a^2 + b^2 + c^2}, &
        \frac{\abs{a}}{a^2 + b^2 + c^2} &\leq \frac{1}{\sqrt{a^2 + b^2 + c^2}}, \\
        \abs{ab} &\leq \frac{1}{2}\left(a^2 + b^2\right), &
        \abs{Ka + Lb} &\leq \max\left\{\abs{K}, \abs{L}\right\} \left(\abs{a} + \abs{b}\right).
    \end{align}
    \end{subequations}
    Using this strategy, we arrive at the following upper bound
    \begin{align}
        \det\left(\mat{I} + \mat{A}_{\bs{\omega}_i}\right) \leq 
        1 - \bigg(& \frac{\abs{u_c}}{u_{\min}} + \frac{\left(\abs{u_c} + \abs{v_c} + \abs{w_c}\right)\left(u_{\min} + \abs{u_c}\right)}{u_{\min}^2} + \frac{\abs{X_{v0}} + \abs{X_{w0}}}{u_{\min}} + 2\abs{X_{v1} - X_{w1} - X_{v1}X_{w1}}\frac{u_{\min}^2 + u_c^2}{u_{\min}^2} \nonumber\\
        &+ \max\left\{\abs{X_{v1}+Z_{w1}+X_{v1}Z_{w1}}, \abs{X_{w1}+Z_{v1}+X_{w1}Z_{v1}}\right\}\frac{u_{\min} + \abs{u_c}}{u_{\min}} 
        + \frac{\abs{X_{v0}X_{w0}}}{u_{\min}^2} + \nonumber\\
        & \max\left\{\abs{X_{v0}}, \abs{X_{w0}}\right\}\frac{u_{\min}^2 + \norm{\mat{V}_c}^2}{u_{\min}^3} 
        + \frac{\abs{X_{v0}}\max\left\{\abs{X_{w1}}, \abs{Z_{w1}}\right\} + \abs{X_{w0}}\max\left\{\abs{X_{v1}}, \abs{Z_{v1}}\right\}}{u_{\min}} \nonumber\\
        & + \abs{X_{v1}\!+\!Z_{w1}\!-\!X_{v1}Z_{w1}}\frac{\abs{v_c}\left(u_{\max} + \abs{u_c}\right)}{u_{\max}^2} + \abs{X_{w1}+Z_{v1}-X_{w1}Z_{v1}}\frac{\abs{w_c}\left(u_{\max} + \abs{u_c}\right)}{u_{\max}^2} \nonumber\\
        & + \frac{\abs{X_{v0}}\left(\abs{X_{w1}u_c}+\abs{Z_{w1}v_c}\right) + \abs{X_{w0}}\left(\abs{X_{v1}u_c}+\abs{Z_{v1}w_c}\right)}{u_{\min}^2} \nonumber\\
        & \triangleq 1 - k_a.
    \end{align}
    Note that the components of the ocean current, $\abs{u_c}$, $\abs{v_c}$, and $\abs{w_c}$, can be upper bounded by $\norm{\mat{V}_c}$.
    We have therefore found a constant upper bound on the determinant.

    Now, let us focus on $\bs{\omega}_{0, i}$.
    Recall the definition of $\bs{\omega}_{0, i}$ in \eqref{eq:omega_v_affine}.
    To find an upper bound, we will use the following inequality
    \begin{align}
        \norm{\mat{v}_i \times \widehat{\bs{\omega}}_{0, i}} &\leq \norm{\mat{v}_i}\norm{\widehat{\bs{\omega}}_{0, i}}, &
        & \implies &
        \norm{\bs{\omega}_{0, i}} &\leq \frac{\norm{\widehat{\bs{\omega}}_{0, i}}}{\norm{\mat{v}_i}}.
    \end{align}
    Recall the definition of $\widehat{\bs{\omega}}_{0, i}$ in \eqref{eq:v_i_dot_matrix_form}.
    To find an upper bound on this vector, we will utilize the following inequality:
    Consider a vector $\mat{x} = \inlinevector{\sum_{i=1}^{N_a} a_i, \sum_{i=1}^{N_b} b_i, \sum_{i=1}^{N_c} c_i}$, where $a_i, b_i, c_i \in \mathbb{R}$.
    The following inequality holds for the Euclidean norm of $\mat{x}$
    \begin{equation}
        \norm{\mat{x}} \leq \sum_{i=1}^{N_a} \abs{a_i} + \sum_{i=1}^{N_b} \abs{b_i} + \sum_{i=1}^{N_c} \abs{c_i}.
    \end{equation}
    Therefore, we can find an upper bound on $\norm{\widehat{\bs{\omega}}_{0, i}}$ by analyzing its components.
    
    Let us begin by investigating the term $\frac{\bs{\upsilon}_{{\rm NSB}, i}\T\dot{\bs{\upsilon}}_{{\rm NSB}, i}}{u_i}$.
    From \eqref{eq:v_NSB_i_nominal}, $\bs{\upsilon}_{{\rm NSB}, i}$ and its time-derivative are given by
    \begin{align}
        \bs{\upsilon}_{{\rm NSB}, i} &= U_{\rm LOS}\mat{R}_p(\xi)\mat{e}_p, &
        \dot{\bs{\upsilon}}_{{\rm NSB}, i} &= U_{\rm LOS}\mat{R}_p(\xi)\left(\bs{\kappa}\times\mat{e}_p + \bs{\iota}\times\mat{p}_{f,i}^f\right).
    \end{align}
    For brevity, let us define
    \begin{equation}
        \mat{e}_d = \bs{\kappa}\times\mat{e}_p + \bs{\iota}\times\mat{p}_{f,i}^f.
    \end{equation}
    Then, the following inequality holds for the investigated term
    \begin{equation}
    \begin{split}
        \abs{\frac{\bs{\upsilon}_{{\rm NSB}, i}\T\dot{\bs{\upsilon}}_{{\rm NSB}, i}}{u_i}} &\leq
        \frac{\norm{\bs{\upsilon}_{{\rm NSB}, i}}\norm{\dot{\bs{\upsilon}}_{{\rm NSB}, i}}}{u_i} =
        \frac{\norm{\mat{v}_i}U_{\rm LOS}\norm{\mat{e}_d}}{u_i} = 
        \norm{\mat{v}_i}\frac{U_{\rm LOS}\norm{\mat{e}_d}\frac{\norm{\mat{e}_p}}{\norm{\mat{e}_p}}}{u_i} = 
        \norm{\mat{v}_i}\frac{\norm{\mat{e}_d}}{\norm{\mat{e}_p}}\frac{\norm{\bs{\upsilon}_{{\rm NSB}, i}}}{u_i} \\
        &\leq \frac{\norm{\mat{e}_d}}{\norm{\mat{e}_p}}\frac{\norm{\mat{v}_i}^2}{u_{\rm min}}
    \end{split}
    \end{equation}
    We can now expand the remaining terms in $\widehat{\bs{\omega}}_{0, i}$ to arrive at the following upper bound
    \begin{equation}
    \begin{split}
        \norm{\widehat{\bs{\omega}}_{0, i}} \leq &
        \frac{\norm{\mat{e}_d}}{\norm{\mat{e}_p}}\frac{\norm{\mat{v}_i}^2}{u_{\rm min}}
        + \abs{\frac{Y_{v1}\left(u_i-u_c\right) + Y_{v0}}{u_i}}v_i^2
        + \abs{\frac{Y_{w1}\left(u_i-u_c\right) + Y_{w0}}{u_i}}w_i^2
        + \abs{\frac{Z_{v0}+Z_{w0}}{u_i}v_iw_i} \\
        &+ \abs{\frac{Y_{v1}u_cv_c - Y_{v0}v_c - Z_{v0}w_c - Y_{v1}u_iv_c}{u_i}v_i}
        + \abs{\frac{Y_{w1}u_cw_c - Y_{w0}w_c - Z_{w0}v_c - Y_{w1}u_iv_c}{u_i}w_i} \\
        &+ \abs{Y_{v0}-Y_{v1}u_c+Y_{v1}u_i}\abs{v_i} + \abs{Z_{v0}w_i} + \abs{Y_{v1}u_cv_c - Z_{v0}w_c - Y_{v0}v_c - Y_{v1}u_iv_c} \\
        &+ \abs{Y_{w0}-Y_{w1}u_c+Y_{w1}u_i}\abs{w_i} + \abs{Z_{w0}v_i} + \abs{Y_{w1}u_cw_c - Z_{w0}v_c - Y_{w0}w_c - Y_{w1}u_iw_c}.
    \end{split}
    \end{equation}
    Next, we use a similar strategy as in the previous section to get the following upper bound
    \begin{equation}
        \begin{split}
            \norm{\widehat{\bs{\omega}}_{0, i}} \leq &
            \frac{\norm{\mat{e}_d}}{\norm{\mat{e}_p}}\frac{\norm{\mat{v}_i}^2}{u_{\rm min}}
            + \left(\max\left\{\frac{\abs{Y_{v1}}\left(u_{\min}+\abs{u_c}\right) + \abs{Y_{v0}}}{u_{\min}}, \frac{\abs{Y_{w1}}\left(u_{\min}+\abs{u_c}\right) + \abs{Y_{w0}}}{u_{\min}}\right\} + \frac{1}{2}\frac{\abs{Z_{v0} + Z_{w0}}}{u_{\min}}\right)\mathrlap{\left(v_i^2 + w_i^2\right)} \\
            &+ \left(\frac{\abs{Y_{v1}u_cv_c} + \abs{Y_{v0}v_c} + \abs{Z_{v0}w_c} + \abs{Y_{v1}u_{\max}v_c}}{u_{\max}} + \abs{Y_{v0}} + \abs{Y_{v1}u_c} + \abs{Z_{w0}}\right)\abs{v_i}
            + \abs{Y_{v1}u_iv_i} \\
            &+ \left(\frac{\abs{Y_{w1}u_cw_c} + \abs{Y_{w0}w_c} + \abs{Z_{w0}v_c} + \abs{Y_{w1}u_{\max}w_c}}{u_{\max}} + \abs{Y_{w0}} + \abs{Y_{w1}u_c} + \abs{Z_{v0}}\right)\abs{w_i}
            + \abs{Y_{w1}u_iw_i} \\
            &+ \left(\abs{Y_{v1}v_c}+\abs{Y_{w1}w_c}\right)\abs{u_i} + \abs{Y_{v1}u_cv_c} + \abs{Z_{v0}w_c} + \abs{Y_{v0}v_c} + \abs{Y_{w1}u_cw_c} + \abs{Z_{w0}v_c} + \abs{Y_{w0}w_c}.
        \end{split}
    \end{equation}
    Note that the norm of $\mat{v}_i$ satisfies
    \begin{equation}
        \norm{\mat{v}_i} = \norm{\bs{\upsilon}_{{\rm NSB}, i}} = U_{\rm LOS}\norm{\mat{e}_p}
        \leq \frac{\norm{\mat{e}_p}}{1 - k_{\rm NSB}}\norm{\mat{v}_u} + \frac{\upsilon_{2, \max} + u_{\min}}{1 - k_{\rm NSB}}\norm{\mat{e}_p},
    \end{equation}
    and the term $\left(v_i^2+w_i^2\right)$ satisfies the following two inequalities
    \begin{align}
        v_i^2+w_i^2 &\leq \norm{\mat{v}_i}^2, &
        v_i^2+w_i^2 &\leq \norm{\mat{v}_u}^2.
    \end{align}
    We finally arrive at the following upper bound on $\norm{\bs{\omega}_{0, i}}$
    \begin{equation}
    \begin{split}
        \norm{\bs{\omega}_{0, i}} \leq &
        \Bigg(\frac{\norm{\mat{e}_d}}{u_{\min}\left(1 - k_{\rm NSB}\right)}
        + \max\left\{\frac{\abs{Y_{v1}}\left(u_{\min}+\abs{u_c}\right) + \abs{Y_{v0}}}{u_{\min}}, \frac{\abs{Y_{w1}}\left(u_{\min}+\abs{u_c}\right) + \abs{Y_{w0}}}{u_{\min}}\right\}
        + \frac{1}{2}\frac{\abs{Z_{v0} + Z_{w0}}}{u_{\min}} \\
        &+ \abs{Y_{v1}} + \abs{Y_{w1}}\Bigg) \norm{\mat{v}_u}
        + \frac{\norm{\mat{e}_d}\left(\upsilon_{2, \max} + u_{\min}\right)}{u_{\min}\left(1 - k_{\rm NSB}\right)}
        + \frac{\abs{Y_{v1}u_cv_c} + \abs{Y_{v0}v_c} + \abs{Z_{v0}w_c} + \abs{Y_{v1}u_{\max}v_c}}{u_{\max}} + \abs{Y_{v0}} \\
        & + \abs{Y_{v1}u_c} + \abs{Z_{w0}} + \frac{\abs{Y_{w1}u_cw_c} + \abs{Y_{w0}w_c} + \abs{Z_{w0}v_c} + \abs{Y_{w1}u_{\max}w_c}}{u_{\max}} + \abs{Y_{w0}} + \abs{Y_{w1}u_c} + \abs{Z_{v0}} \\
        & + \abs{Y_{v1}v_c} + \abs{Y_{w1}w_c} + 
        \frac{\abs{Y_{v1}u_cv_c} + \abs{Z_{v0}w_c} + \abs{Y_{v0}v_c} + \abs{Y_{w1}u_cw_c} + \abs{Z_{w0}v_c} + \abs{Y_{w0}w_c}}{u_{\min}} \\
        & \triangleq a_v\norm{\mat{v}_u} + b_v
    \end{split}
    \end{equation}
    Similarly to the previous section, we can upper-bound $\abs{u_c}$, $\abs{v_c}$, and $\abs{w_c}$ with $\norm{\mat{V}_c}$.
    We have thus found positive constants $a_v$ and $b_v$ that satisfy \eqref{eq:omega_0_bound}.

\end{document}